\documentclass[
showpacs,
 amsmath,amssymb,
 aps,
 10pt,
twocolumn
]{revtex4-1}

\usepackage[usenames,dvipsnames]{color}
\usepackage{graphicx,textcase} 
\usepackage{dcolumn,overpic} 
\usepackage{bm,color}
\usepackage[T1]{fontenc}
\usepackage{ae,aecompl}
\usepackage{chngcntr}
\usepackage[para]{threeparttable}
\usepackage{etoolbox}
\usepackage{lipsum}
\usepackage[bottom]{footmisc}
\usepackage{setspace}
\AtEndEnvironment{table*}{\vskip10pt}{}{}

\begin{document}

\title{Exchange interactions of CaMnO$_3$ in the bulk and at the surface}
\author{S.~Keshavarz$^1$, Y.~O.~Kvashnin$^1$, D.~C.~M.~Rodrigues$^{1,2}$, M.~Pereiro$^1$, I.~Di~Marco$^1$, C.~Autieri$^1$, L. Nordstr\"om$^1$, I.~V.~Solovyev$^{3,4}$, B.~Sanyal$^1$, and O.~Eriksson$^1$}

\affiliation{$^1$Uppsala University, Department of Physics and Astronomy, Division of Materials Theory, Box 516, SE-751 20 Uppsala, Sweden}
\affiliation{$^2$Faculdade de F\'isica, Universidade Federal do Par\'a, CEP 66075-110, Bel\'em, Par\'a, Brazil}
\affiliation{$^3$National Institute for Materials Science, 1-1 Namiki, Tsukuba, Ibaraki 305-0044, Japan}
\affiliation{$^4$Department of Theoretical Physics and Applied Mathematics, Ural Federal University, Mira Street 19, 620002 Ekaterinburg, Russia}

\date{\today}

\begin{abstract}
In this work, we present electronic and magnetic properties of CaMnO$_3$ (CMO) as obtained from \textit{ab initio} calculations.
We identify the preferable magnetic order by means of density functional theory plus Hubbard $U$ calculations and extract the effective exchange parameters ($J_{ij}$'s) using the magnetic force theorem.
We find that the effects of geometrical relaxation at the surface as well as the change of crystal field are very strong and are able to influence the lower energy magnetic configuration.
In particular, our analysis reveals that the exchange interaction between the Mn atoms belonging to the surface and the subsurface layers is very sensitive to the structural changes.
An earlier study [A. Filippetti and W.E. Pickett, Phys. Rev. Lett. \textbf{83}, 4184 (1999)] suggested that this coupling is ferromagnetic and gives rise to the spin flip process on the surface of CMO.
In our work we confirm their finding for an unrelaxed geometry, but once the structural relaxations are taken into account, this exchange coupling changes its sign.
Thus, we suggest that the surface of CMO should have the same $G$-type antiferromagnetic order as in the bulk.
Finally, we show that the suggested SF can be induced in the system by introducing an excess of electrons.
\end{abstract}


\maketitle

\section*{Introduction}

Magnetic transition metal (TM) perovskites are an extremely fascinating class of materials, exhibiting a sophisticated interplay between structure, charge, orbital and spin degrees of freedom~\cite{dagotto-book}.
As a result, depending on their composition, they exhibit a variety of different ordered states and possess various useful physical properties, such as multiferroicity~\cite{BS-and-OE-book,eric}, non-collinear magnetism~\cite{TbMnO3-Mostovoy}, half-metalicity and colossal magnetoresistance (CMR) [see e.g. Ref.~\cite{CMR-LCMO}], which make these materials attractive for spintronic applications.
Nowadays very clean surfaces and sharp interfaces between different perovskites can be synthesised with great precision.
Exotic phenomena such as superconductivity~\cite{lao-sto-supercond} and a realization of 2D electron gas~\cite{2deg,surf-2deg} have been reported for 2D-derived perovskite materials.
This technological advancement boosted not only the experimental investigations in this direction, but also lead to many fundamental questions for theory. 

One of the fundamental questions for magnetic TM perovskites is: how does a system react microscopically when it is confined to two dimensions?
Lowering of the symmetry and a reduced coordination number can drastically change its electronic structure.
Due to the above-mentioned coexistence of several degrees of freedom, these changes are difficult to predict in TM perovskites and first-principle electronic structure calculations are necessary. 
Among perovskite systems, the mixed valence manganites are of particular interest~\cite{CMR-LCMO,lcmo-phasediag,cmo-polarons,silvia}.

The present study concerns a classical TM oxide: CaMnO$_3$ (CMO). 
In bulk, the sixfold coordinated Mn ions form a set of half-filled $t_{2g}$ orbitals, that corresponds to the formal valence state Mn$^{4+}$.
Strongly antiferromagnetic (AFM) superexchange interactions~\cite{pwa-superexch} between these ions give rise to a G-type AFM order, which is stabilized below the N\'eel point of 120 K~\cite{optical-gap,distortion}.

However, at the surface of CMO, the Mn ions are surrounded by only 5 oxygen atoms. This may lead to a dramatic reconstruction of the electronic structure, when the less bonded Mn $e_g$ states shift to the low-energy region and form the metallic bands at the Fermi level. In this case, the the ferromagnetic (FM) double-exchange mechanism~\cite{zener-de,double-1,double-2} also comes into play and will
compete with the superexchange coupling.
Hence the exchange interactions ($J_{ij}$'s) at the surface of CMO will be defined by the subtle balance between these two contributions, and are therefore expected to be very different from the bulk values.

Filippetti and Pickett~\cite{pickett-1} investigated the surface of CMO by means of density functional theory (DFT).
The authors considered a slab of 6-layer thickness, analysed its band structure and calculated several magnetic configurations.
They found that the surface and subsurface Mn spins are coupled ferromagnetically through a double-exchange-type mechanism.
This change of sign of the exchange interaction results into a magnetic reconstruction at the surface of CMO, where the surface spins reverse their orientations, i.e. undergo a spin-flip (SF) process.

In the present study we reexamine the electronic and magnetic structure of the surface of CMO in presence of strong on-site correlation effects between $3d$ electrons. 
The calculations are based on a combination of DFT plus a static mean-field approach (DFT+$U$)~\cite{stoner}.
The preferable magnetic order is identified by means of two complementary approaches: direct total energy comparison and extraction of effective $J_{ij}$'s by means of the magnetic force theorem.
We also point out the importance of structural relaxation effects at the surface and their influence on the exchange interactions. 

This paper is organized as follows. 
In section~\ref{theory} we briefly explain the computational details and the methods used in this work. 
Section~\ref{results} contains the main results of this study and is divided in several subsections. 
First the magnetic properties of bulk CMO are presented, next we discuss the results of the slab calculation with and without geometrical relaxation.
For all considered cases we identify the preferable magnetic order and analyze in detail the $J_{ij}$'s in the system.
We also discuss the possibility of modifying the magnetic order on the surface of CMO by adding a uniform electron doping.
Finally, we draw our conclusions to this work in section~\ref{conclusions}.

\section{Computational Details}
\label{theory}

We start by presenting the structural model used in this work.
Bulk CMO crystallizes in the orthorhombic perovskite structure having \textit{Pnma} space group with the lattice parameters $a$= 5.28~\AA, b=0.99a and c=$\sqrt{2}a$.~\cite{distortion} 
In this structure the Mn-O-Mn angle is 158$^{\circ}$ rather than 180$^{\circ}$ as in the ideal perovskite structure. 
This structure was taken as the starting point for all the calculations performed in this work.
For the surface simulations, we have used a supercell geometry consisting of 6 alternating layers of CaO mediated and terminated by 7 layers of MnO$_2$ stacked along (001) direction. 
Since the periodic boundary conditions are used in all of the three dimensions, a 20-\AA-thick layer of vacuum is used in the construction of the supercell. 

The relaxation of the bulk and the slab geometries has been performed on the entire structure (i.e. both forces and the volume were optimized) for different magnetic orders and computational setups (with and without Hubbard $U$). 
The optimal geometries were obtained using the projector augmented wave method~\cite{paw}, as implemented in the VASP code~\cite{vasp}.
The \textit{k} integration over the Brillouin zone has been performed using 7$\times$7$\times$5 points for bulk and 7$\times$7$\times$1 points for the slabs. 
Plane-wave energy cut-off was set to 550 eV. We used an exchange-correlation functional based on local spin density approximation (LSDA)~\cite{lda}, following Refs.~\cite{pickett-1,pickett-2}
More details on the geometry of the bulk and the slab after relaxation will be given in the next sections. 

As mentioned before, in order to improve the description of relatively localized TM $3d$ electrons, the LSDA+$U$ technique has been applied. 
For $3d$ orbitals, where the electrons are expected to find more atomic-like characters, the Coulomb interaction matrix can be formulated via Slater integrals $F^n$~\cite{stoner}. The Coulomb parameters can be defined as follows 
\begin{equation}
U = F^0, \qquad J=\frac{F^2+F^4}{14},
\end{equation}
where $U$ is Hubbard parameter and $J$ is Hund's exchange.
Conversely from $U$ and $J$ one can recover the values of the Slater integrals by assuming the atomic like ratio F$^2/$F$^4$= 0.625.
The values of $U$ and $J$ can be either extracted from experiments or from first principles calculations. 
To our best knowledge, there is a lack of experimentally available U value for this system. 
Therefore, we rely on the theoretically calculated values. In the work by Hung~\cite{u-j}, the $U_{eff}=U-J$ value is extracted by fitting energy differences obtained from LSDA+$U$ for different $U$ values to the one obtained from hybrid-functional energy, for the same electronic structure. In this way, the effective $U$ is obtained to be about 3 eV. Therefore, we used 4 eV for the $U$ and 0.9 for the $J$, following the choices of the parameters used in prior studies~\cite{optic,gap-1}.

For the choice of the double-counting (DC), we have used the fully localized limit (FLL) formulation~\cite{FLL-DC} both for bulk and surface. 
This DC is usually appropriate for the systems wherein the electrons are close to the atomic limit, e.g., TM oxides. 
However, additional simulations with the around mean field (AMF) DC~\cite{amf-2} were performed and were found to qualitatively confirm the FLL results. 
Therefore, to avoid presenting two sets of qualitatively similar results, we only show results obtained with the FLL DC.

Once the atomic positions  in the ground state have been identified, the electronic structure and the magnetic properties of the CMO bulk and slabs are investigated in the framework of a full-potential linear muffin-tin orbital (FP-LMTO) code RSPt~\cite{rspt}, in the scalar relativistic approximation. 
The full potential character of this code enables us to study non closed-packed structures, e.g. surfaces. 
The details of this implementation, including dynamical mean field theory, can be found in Ref.~\cite{rspt,igor,oscar,igor-2}. 

Once the electronic structure has been converged, the next step is to calculate the exchange parameters. 
This is achieved by mapping the magnetic excitations onto the Heisenberg Hamiltonian:
\begin{eqnarray}
\hat H = -\sum_{i \ne j} J_{ij} \vec e_i\cdot \vec e_j ,
\end{eqnarray}
where $J_{ij}$ is an exchange interaction between the two spins, located at sites $i$ and $j$, and $\vec e_i$ is a unit vector along the magnetization direction at the corresponding site.
The exchange parameters can be extracted in several ways, e.g., frozen magnon approximation~\cite{fma} or the Liechtenstein-Katsnelson-Antropov-Gubanov method (LKAG)~\cite{liechten1, liechten2}. The former works in the reciprocal space and the latter, in contrast, in the real space. However, one should notice that these two methods are basically equivalent and their main quantities are related to each other by Fourier transform. In this work, we have obtained the pair-wise exchange interactions using the latter method which is based on the energy difference between an infinitesimal rotation of the spins in the system in the limit of local force theorem. 
More specific details about the evaluation of the exchange parameters, especially with respect to the choice of the basis set used for localized orbitals, can be found in Ref.~\cite{Jijs-in-rspt}. Finally, in order to calculate the ordering temperature for the bulk of CMO based on the extracted exchange parameters, we used classical Monte Carlo method as implemented in the UppASD code~\cite{uppasd}. 

We would like to mention that spin-orbit (SO) coupling has not been taken into account in this work.
The SO effects are mainly responsible for other types of magnetic interactions, e.g., anisotropic exchange interactions and magnetocrystalline anisotropy. 
However, the bilinear Heisenberg term, considered in the present work, is usually the leading one primarily determining low-temperature magnetic order. 
Nevertheless, we performed additional simulations, which suggested that SO interaction does not affect the conclusions of our work. 

\begin{figure*}[tp]   
a) \includegraphics[width=0.3\textwidth]{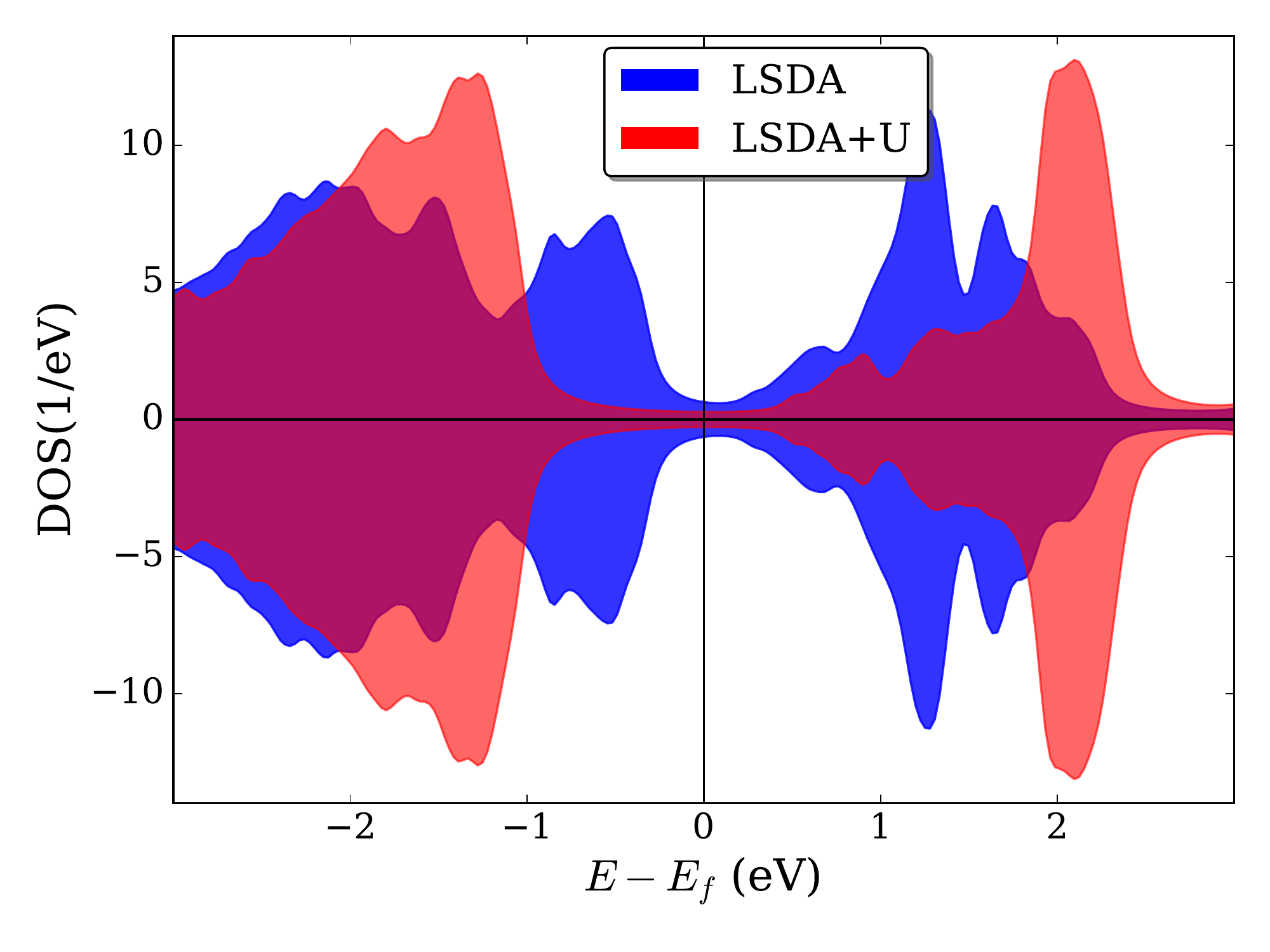}
b) \includegraphics[width=0.3\textwidth]{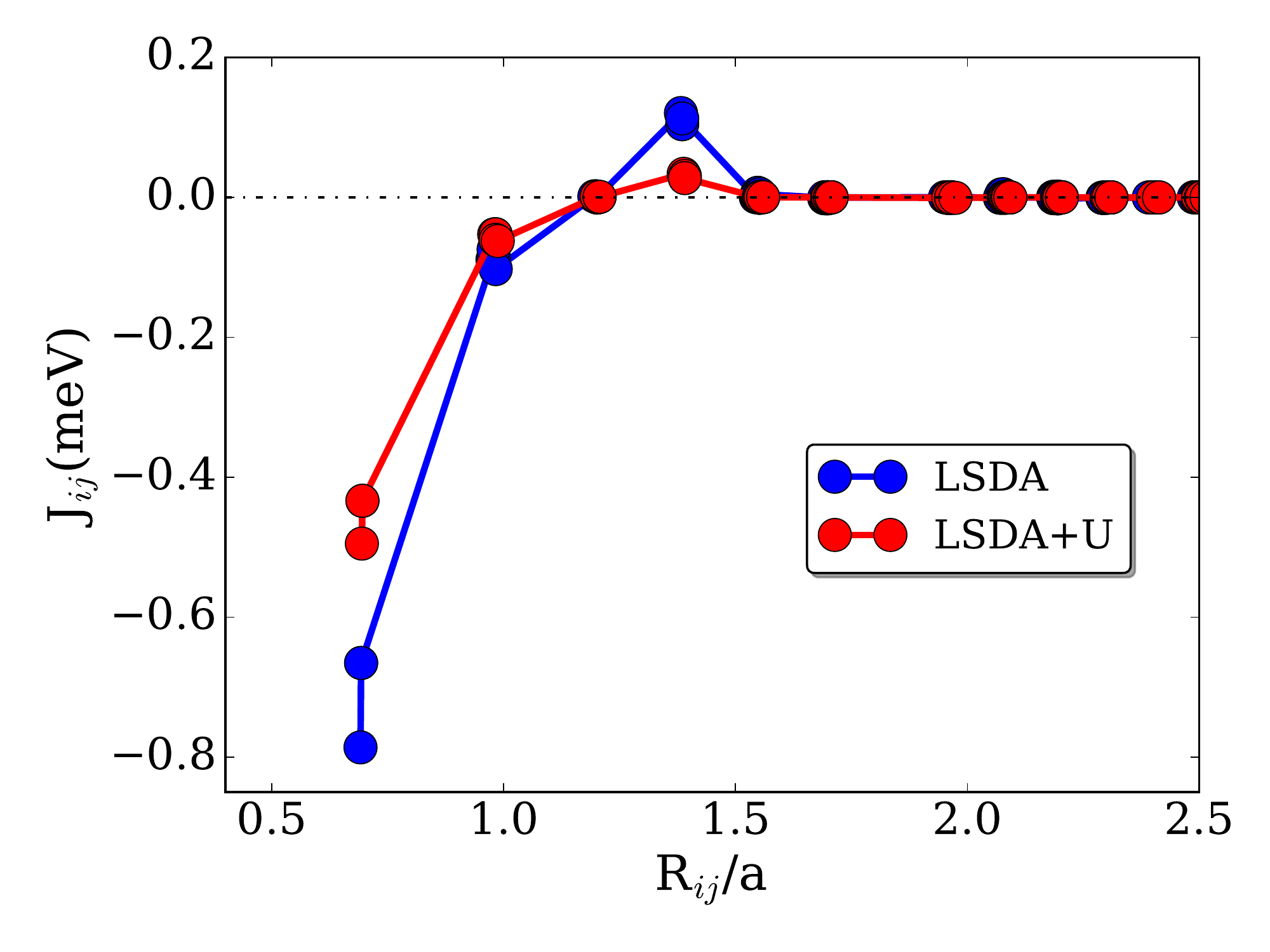}
c) \includegraphics[width=0.3\textwidth]{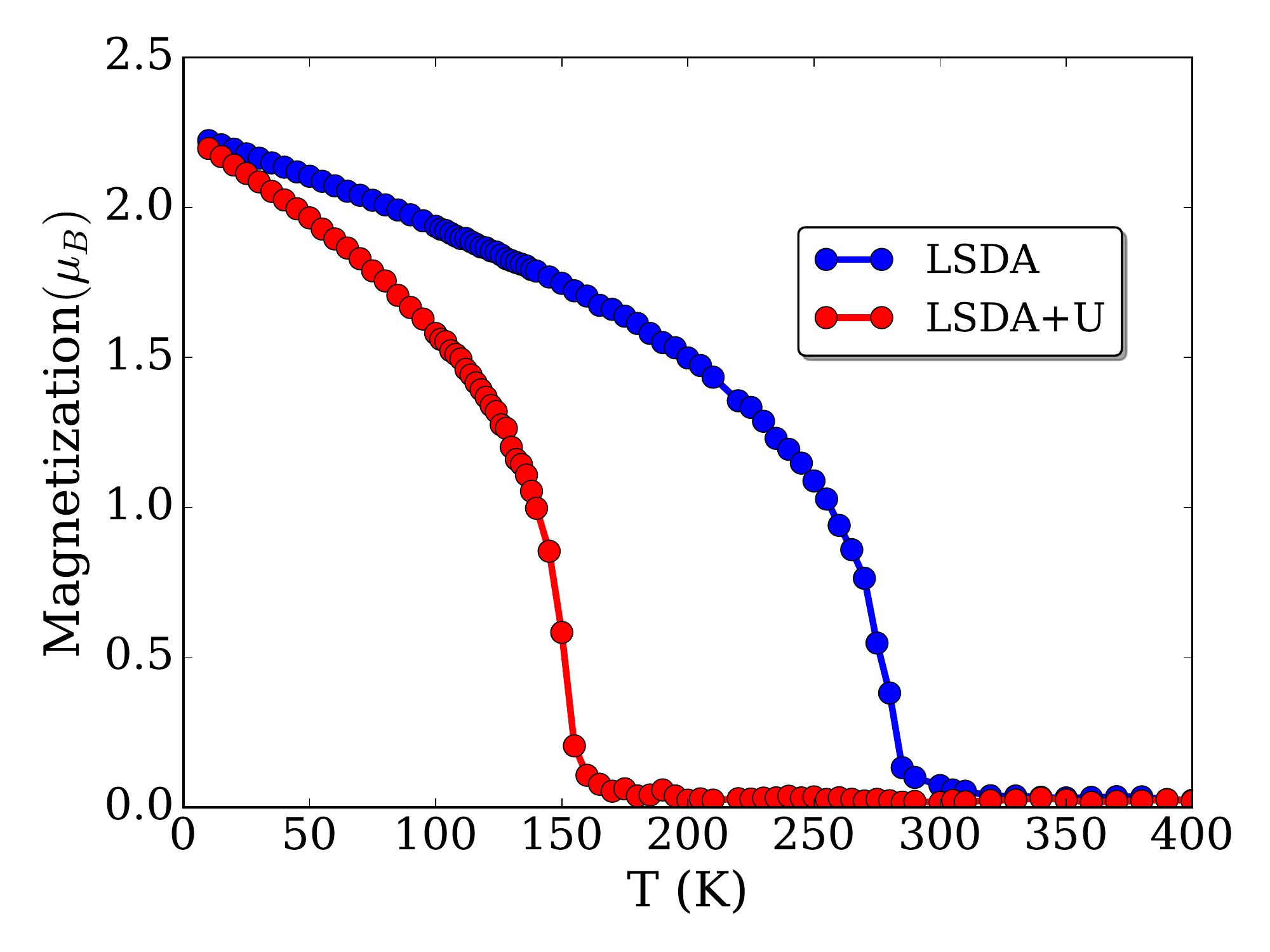}
 \caption{Information for the bulk of CMO. a) the total density of states of the majority- and minority-spin components in LSDA and LSDA+$U$ methods. b) The exchange interaction between the Mn atoms as well as c) the projected magnetic moments of Mn atoms in LSDA (blue line) and in LSDA+$U$ (red line).}
\label{fig:bulk}
\end{figure*}

\section{Results and discussions}
\label{results}
\subsection{Bulk of CaMnO$_3$}
\label{bulk-of-cmo}
As experimentally and theoretically reported, CMO in bulk form is a G-type antiferromagnet~\cite{gtype-1,gtype-2}. This means that both intra- and inter-plane coupling between the Mn atoms are antiferromagnetic. Based on this knowledge, we performed our simulations for bulk considering only G-type AFM order. 
In Fig.~\ref{fig:bulk}, the total density of states (DOS) for the relaxed structure is presented. As seen from the plot, the bulk properties obtained from LSDA and LSDA+$U$ are substantially different, especially when it comes to the heights and the positions of the peaks. 
This is so, because LSDA+$U$ has a tendency to increase the band gap of LSDA by shifting the $3d$ states away from the Fermi level ($E_F$). 
Due to the distortion of the MnO$_6$ octahedra, the cubic degeneracies of the \textit{d} orbitals are lifted. However, the splitting between the states is small and, therefore, we can still separate the states into $e_g$-like and $t_{2g}$-like states. 
With this definition in mind, we observe that the gap is opened between the $t_{2g}$ $(d_{xz},d_{xy},d_{yz})$ states below the Fermi level and $e_g$ ($d_{xy},d_{z^2}$) states above the Fermi level. Therefore, this material can be classified as a Mott insulator according to the Zaanen-Sawatzky-Allen diagram~\cite{mott-ins}. 
The value of the band gap is found to be 0.4 eV and 1.1 eV in LSDA and LSDA+$U$, respectively. 
The band gap in LSDA is in close agreement with the results from earlier electronic structure calculations~\cite{pickett-1,nguyen}, while the band gap in LSDA+$U$ is in a fair agreement with the experimental optical band gap (1.55 eV) according to Ref.~\cite{optical-gap}. Moreover, in the work by Jung \textit{et al.}~\cite{optic}, the band gap is found to be around 1 eV based on the optical-conductivity analyses. However, in that study the opening of the band gap arises between the O 2\textit{p} states (below the Fermi level) and Mn \textit{d} states (above). This corresponds to a charge transfer insulator in the Zaanen-Sawatzky-Allen diagram.

\begin{table}
\caption{\label{tab:table-bulk} Calculated spin moment of Mn atoms (in $\mu_B$) as well as the total and and the orbital-decomposed $J_{ij}$'s (in mRy) between two closest neighboring atoms in the bulk of CMO.}
\begin{ruledtabular}
\begin{tabular}{lcccccc}
  & M$_s$ & Total &  $e_g-e_g$  &   $t_{2g}-t_{2g}$  &    $e_g-t_{2g}$ \\
\hline
LSDA& 2.31 \:\: & -0.786 & 0.018 & -1.096 & 0.291 \\
LSDA+$U$& 2.55 \:\: & -0.495 & -0.002 & -0.646 & 0.153 \\
LSDA-ASW\footnotemark[1]& 2.44 \:\: & -0.704 \\
LSDA\footnotemark[2] & 2.36 \:\: & -1.911 & \\
Exp.\footnotemark[3] &   & -0.485 & \\
\end{tabular}
\begin{tablenotes}
\item[a] Ref.~\cite{cardoso}, with the reported ordering temperature of 434 K based on mean field approximation. \: \: \:  \: \\
\item[b] Ref.~\cite{pickett-1} \qquad \: \qquad \qquad \qquad \: \: \: \qquad \qquad \: \: \quad \: \: \: \: \quad \: 
\item[c] Ref.~\cite{j-exp} \qquad \: \qquad \qquad \qquad \: \: \: \qquad \qquad \: \quad \: \: \: \: \: \quad \:
\end{tablenotes} 
\end{ruledtabular}
\end{table}

Further insides into the magnetic structure can be obtained by calculating the exchange parameters ($J_{ij}$). Panel b of Fig.~\ref{fig:bulk} illustrates the results for $J_{ij}$ in which the negative values indicate AFM couplings. The points in the graph located around $R_{ij}/a=1/\sqrt{2}$ are considered to be the first nearest neighbors (NN), while the second nearest neighbor is the one located close to $R_{ij}/a=1.0$. 
Fig.~\ref{fig:bulk} b shows two nearest neighbor exchange interactions, in which the differences are due to the structural distortion in the $xy$ plane, which makes $x$ and $y$ axes inequivalent.
The exchange parameters with the first and the second nearest neighbors are found to be AFM, which confirms the G-type AFM order for this system as the ground state in both LSDA and LSDA+$U$ methods. 
Comparing these values with the ones obtained for the unrelaxed structure (in which the experimental lattice constant is used), we find a few percent enhancement in $J_{ij}$'s which is due to the small reduction in the lattice constant of LSDA (5.19\AA) and LSDA+$U$ (5.22\AA) versus the experimental lattice constant (5.28\AA). This reduction, in turn, leads to the tiny change in the angle of $\widehat{Mn-O-Mn}$ bond (156.3$^{\circ}$) compared to the unrelaxed structure (156.8$^{\circ}$) in both methods, and a small reduction of 2.5\% in the spin moment of each Mn atom. All these changes together modify the super-exchange interaction between Mn atoms, which results in a minor modification of $J_{ij}$ values.

Further, one can notice that for bulk CMO, the $J_{ij}$'s extracted from LSDA+$U$ are relatively suppressed as compared to those of LSDA. 
This can be understood from the fact, that the exchange interaction in TM oxides is proportional to $t^2/\Delta_{s.f.}$ (see e.g.~\cite{mazin}), where $t$ is the effective inter-site hopping integral and $\Delta_{s.f.}$ is the energy cost of a spin flip.
The latter is equal to Stoner $I$ in LSDA. However, in LSDA+$U$ the U parameter can explicitly contribute to the intraatomic spin splitting~\cite{Hubbard-Stoner}.
Since $U$ is usually larger than $I$, and also because of the reduction of $t$ value caused by the localization, an overall suppression of the $J_{ij}$ can be anticipated, and is also found here (see Table~\ref{tab:table-bulk}). The obtained value of $J_{ij}$ in the framework of LSDA+$U$ is in a good agreement to the experimental value of 0.485 mRy~\cite{j-exp}.

Keeping in mind that in an insulating material like CMO, the dominating exchange mechanism is basically the super-exchange, one can speculate that the predicted exchange coupling within LSDA is rather overestimated. 
This can be seen by comparing the ordering temperature (T$_N$) obtained using the LSDA and LSDA+$U$ derived exchange parameters. Using classical MC simulations as implemented in the UppASD code, we obtained the ordering temperature of 285 and 150 K in LSDA and LSDA+$U$, respectively. The ordering temperature is considered where the sublattice magnetic moment approaches 0, as is shown in panel c of Fig.~\ref{fig:bulk}. A residual magnetization is expected after the critical temperature due to finite size effect in the simulations.
The experimental ordering temperature is around 120 K~\cite {optical-gap,distortion}. 
Hence, LSDA+$U$ based exchange parameters describe more accurately the ordering temperature of CMO, which demonstrates that on-site Coulomb repulsion among the $3d$ orbitals of Mn plays an important role.

Finally, in order to have a further insight to the mechanism of the exchange coupling in the system, in Table~\ref{tab:table-bulk} we present the spin moment as well as the total and orbital decomposed $J_{ij}$'s between the nearest Mn atoms obtained within LSDA and LSDA+$U$. One can see that, the $e_g$-derived contributions are practically negligible, since these states are almost empty and therefore they can not participate in the exchange interactions. 
On the contrary, the $t_{2g}-t_{2g}$ contribution is the decisive component for the general magnetic behavior of the system and gives rise to the G-type AFM order.
There is also a small $e_g-t_{2g}$ term, which appears due to a not perfect relative alignment of the neighboring MnO$_6$ octahedra.

\subsection{Results for the slab}

\begin{figure}[tp]   
 \includegraphics[width=0.36\textwidth]{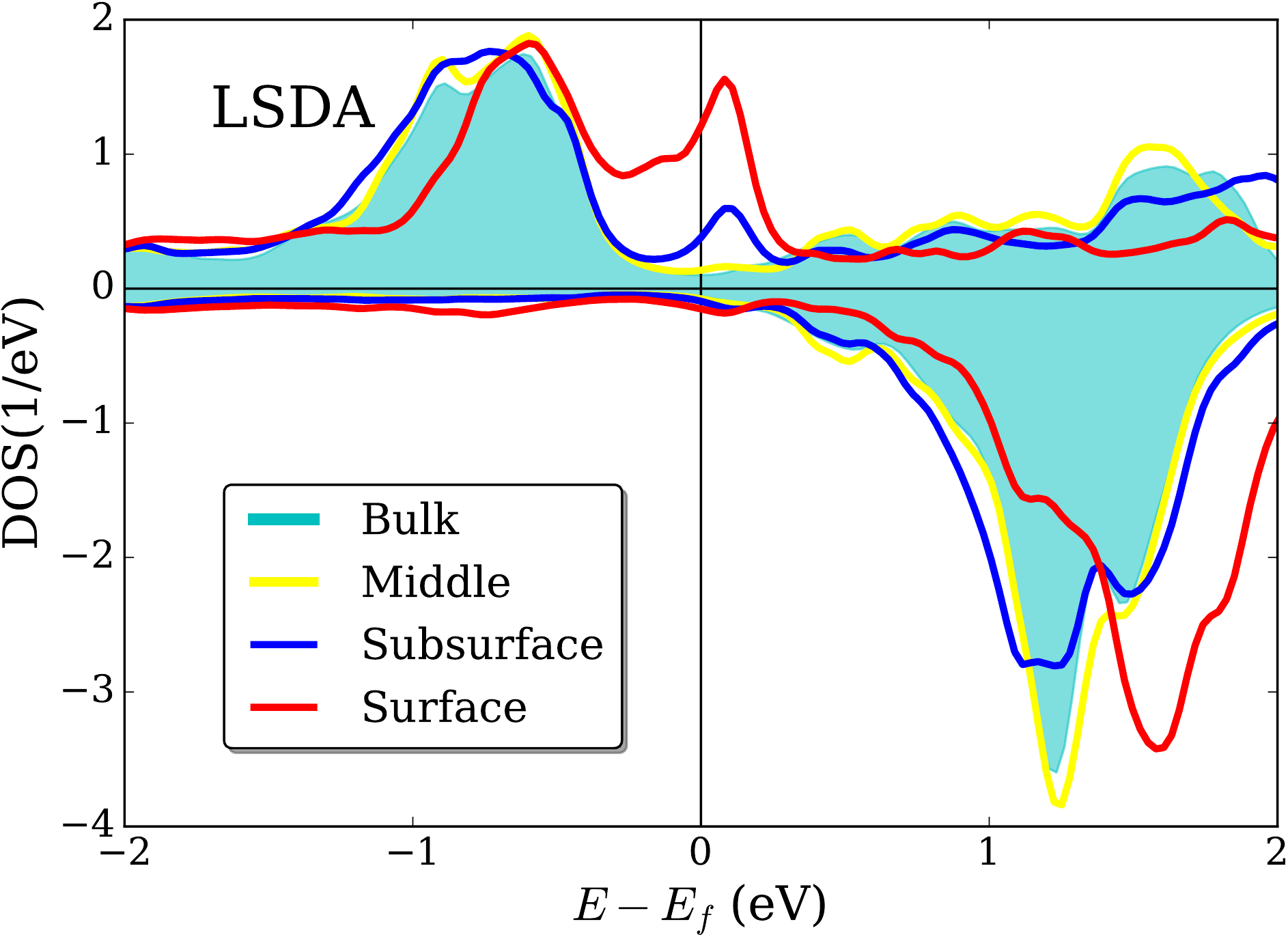}
 \includegraphics[width=0.36\textwidth]{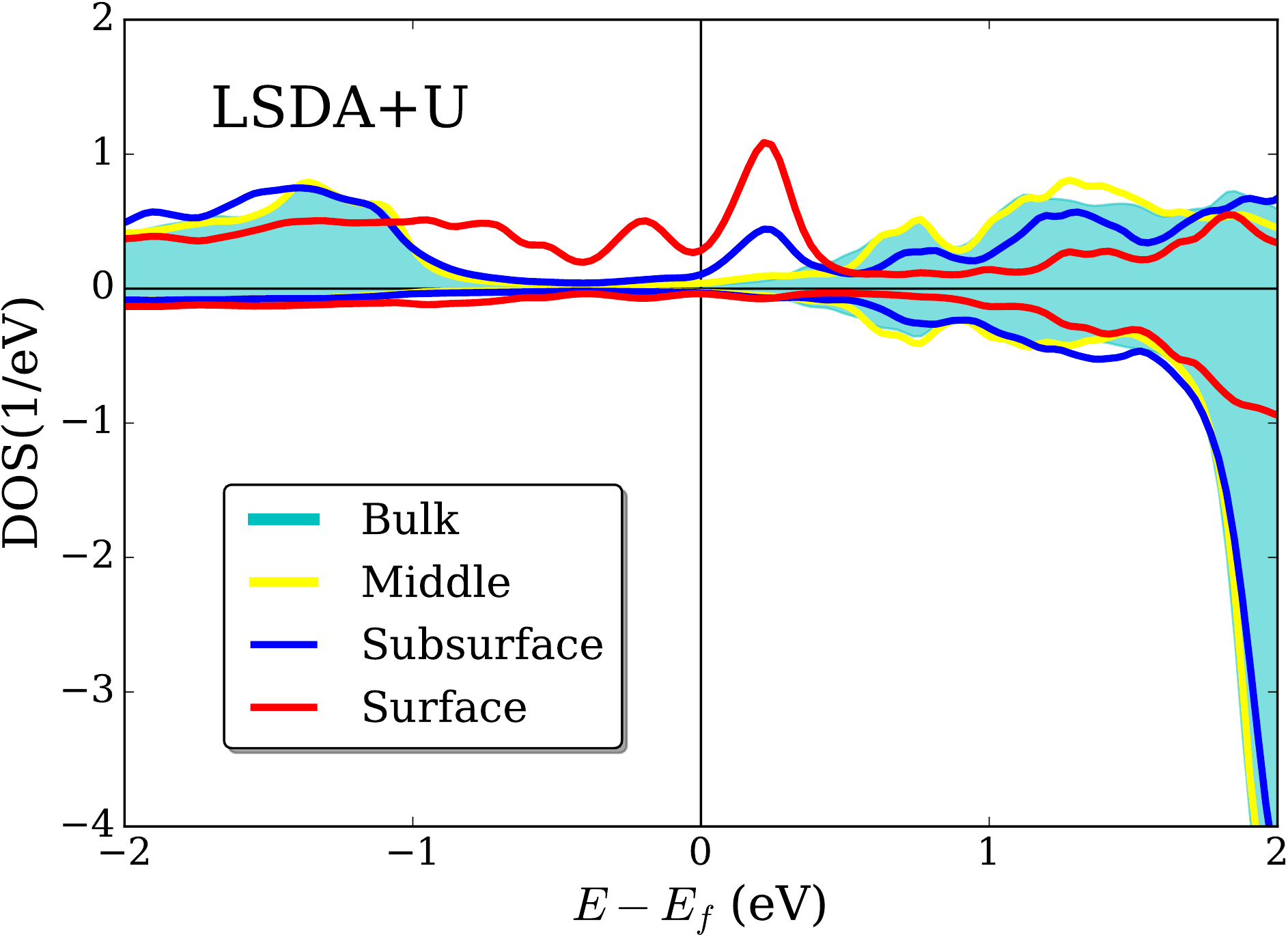}
 \caption{The total density of states of the unrelaxed structure of CMO slab for the majority- and minority-spin components in LSDA (top panel) and LSDA+$U$ (bottom panel).}
\label{fig:slab-dos-unrel}
%
 \includegraphics[width=0.4\textwidth]{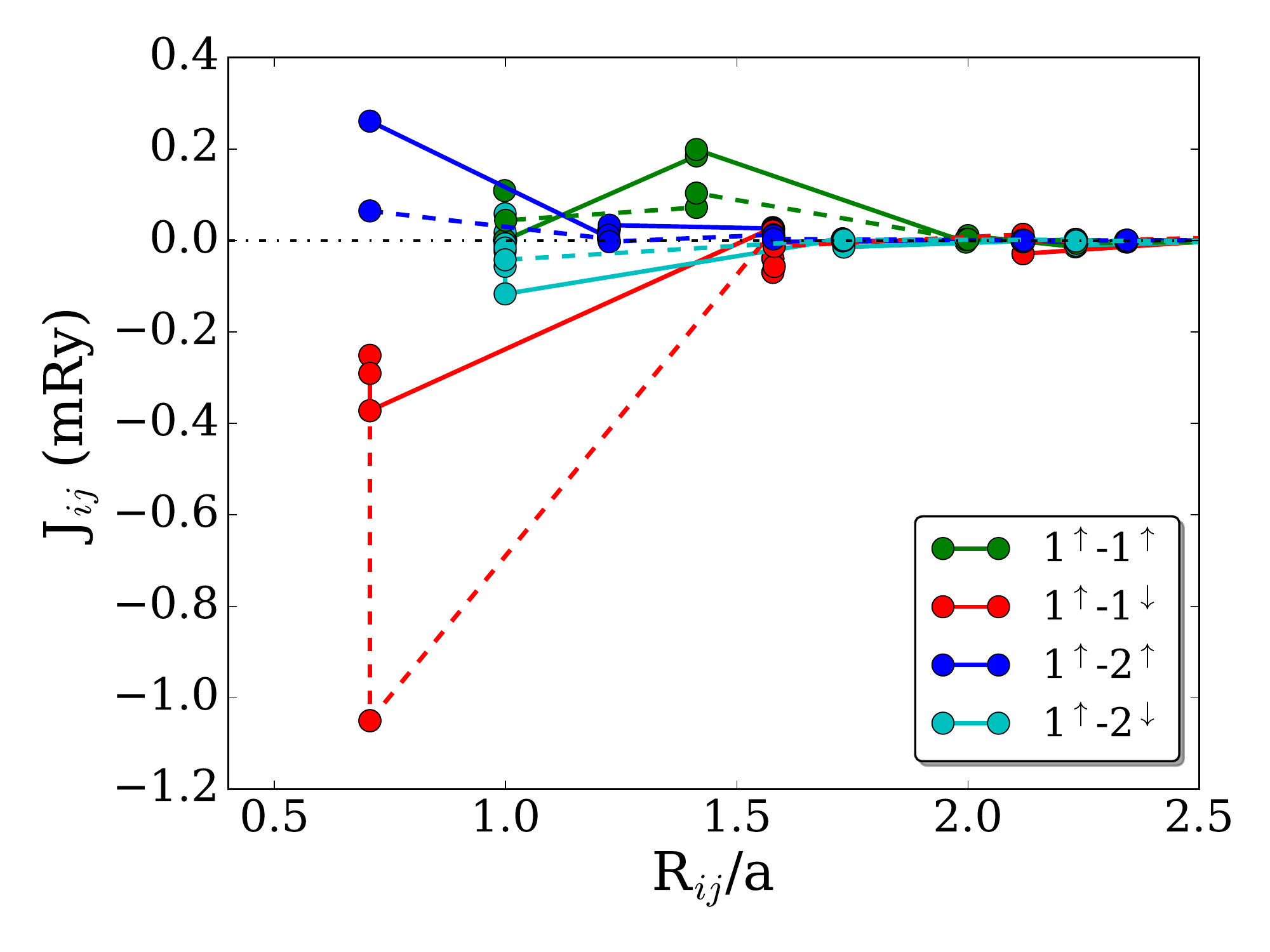}
 \includegraphics[width=0.21\textwidth]{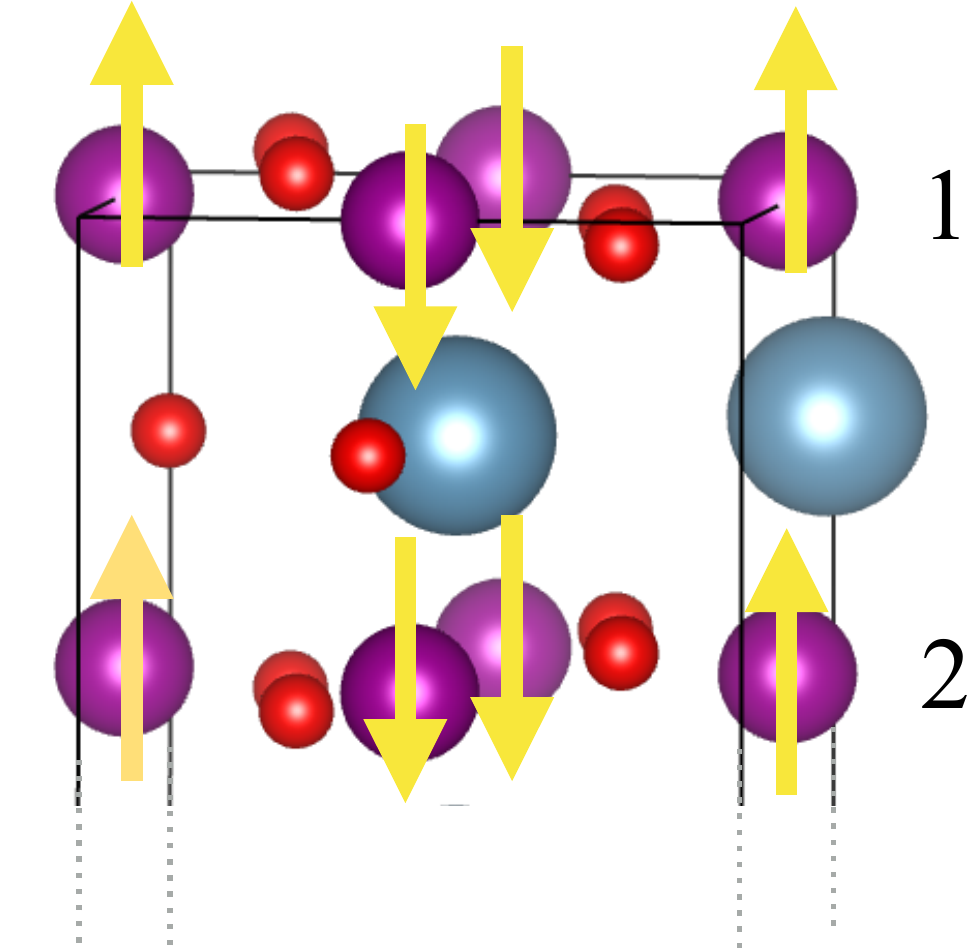}
 \caption{Top panel: The exchange interaction between an atom at the surface of CMO for an unrelaxed slab with the atoms at the surface (layer 1) and subsurface (layer 2) as a function of distance, from calculations based on LSDA (solid lines) and LSDA+$U$ (dashed lines). Bottom panel: The obtained ground state magnetic configuration for the unrelaxed surface of CMO. The purple balls represent Mn atoms with yellow arrow indicating spin magnetization direction, the red balls are O and the blue ones are Ca atoms.}
\label{fig:slab-j-unrel}
\end{figure}

\subsubsection{Unrelaxed slab} 

We start by considering an unrelaxed free-standing slab of CMO as a truncated bulk structure. As mentioned in Section~\ref{theory}, the slab consists of 6 alternating layers of CaO mediated and terminated by the layers of MnO$_2$ from both sides, as shown in the bottom panel of Fig.~\ref{fig:slab-j-unrel}. This thickness for the slab has been chosen by comparing the spin moment of the Mn atom in the innermost layer and in the bulk, where this difference is obtained to be less than 0.2\%. In addition, the comparison of the projected density of states (PDOS) of the Mn \textit{d} orbitals of the innermost layer with the PDOS of the bulk does not show any visible differences on the scale of interest, as can be seen in Fig.~\ref{fig:slab-dos-unrel}.

In this work, the exchange parameters are calculated for different layers as a function of the interatomic distance between the interacting atoms. The aim is to understand in detail the differences between bulk and surface, and also to see how the additional correction in LSDA+$U$ would affect the exchange interactions. First of all, we focus on presenting the electronic structure results especially for the density of states which have decisive roles in determining the magnetic behavior of the system.
In Fig.~\ref{fig:slab-dos-unrel}, the PDOS of the surface, subsurface and the innermost (middle) layers of the unrelaxed CMO slab as well as that of the bulk are presented for both LSDA and LSDA+$U$ methods. 
As can be seen, the spectrum of the middle layer mimics rather well the spectrum of the bulk. A tiny difference between these two spectra can be related to the difficulties in matching the same spacial points for the sampling of a two-dimensional and three-dimensional Brilluin zone.

According to the LSDA calculations (top panel of Fig.~\ref{fig:slab-dos-unrel}), the surface and the subsurface of the slab turn to be metallic, while the middle layer stays insulating, similar to the bulk. This is due to the broken symmetry at the surface which lifts the degeneracy of the $e_g$ states and results in a metallic character for surface and subsurface. This metallic character mainly comes from the presence of the $d_{z^2}$ states at $E_f$. The results of LSDA+$U$, to some extent, confirm the LSDA predictions except that for the subsurface the metallicity is almost quenched, although not completely. However, since LSDA+$U$ have a large tendency to highlight the atomic features, the predicted band gap for the innermost layer and the subsurface are substantially larger than the corresponding ones obtained in LSDA.

In the work reported by Pickett~\cite{pickett-1}, the ground state for the slab of CMO is observed to be a G-type AFM, except that for the surface atoms the spin moments are flipped such that each atom at the surface is coupled with the atom in the subsurface ferromagnetically (see bottom panel of Fig.~\ref{fig:slab-j-unrel}). In other words, a SF occurs for the surface Mn atoms. This interesting possibility motivated us to check the electronic structure of the CMO slab for the G-type AFM with and without a SF at the surface. For simplicity, we refer the former as AFM and the later as the SF configuration.
Comparing the total energies of these states, our results for the unrelaxed slab completely confirm the findings in Ref.~\cite{pickett-1}, with 78(22) meV lower total energy for the SF in LSDA(LSDA+$U$).
 Metallicity for the surface atoms is predicted in both LSDA and LSDA+$U$ results (see Fig.~\ref{fig:slab-dos-unrel}). 

We proceed now to an analysis of the interatomic exchange couplings ($J_{ij}$). 
Using the SF magnetic order as the reference state, the exchange parameters between the atoms at the surface and the subsurface are calculated. The results are shown in Fig.~\ref{fig:slab-j-unrel}, wherein surface atoms are labeled as 1 and subsurface atoms as 2. The exchange parameters between the atoms in the inner layers are not shown since they are quite similar to the corresponding results for the bulk illustrated in panel b of Fig.~\ref{fig:bulk}. The most interesting observation is that the coupling of the atom at the surface and the one right below at the subsurface is FM in both LSDA and LSDA+$U$ (see the blue lines). In addition, the AFM coupling between the closest Mn atoms at the surface is also observed (red lines). Although the latter is substantially weaker than that in the bulk.
These two observations confirm the SF scenario at the surface, which was obtained in Ref. ~\cite{pickett-1}, by total energy comparison. 

One has to always remember that the $J_{ij}$'s extracted by means of the LKAG method usually depend on the magnetic configuration they are extracted from. 
This dependence simply arises because the electronic structure in various magnetic states can be substantially different.
Therefore, it is important to verify that the reference state is the actual ground state, otherwise there is no guarantee that the obtained results can predict the correct magnetic properties such as ordering temperature, magnon spectra, etc. 
We have also calculated the $J_{ij}$'s from the G-type AFM order without the surface SF and these results indicated an instability of this state, suggesting that the SF scenario is the ground state.
Thus, we can claim that the results provided by all different methods offer a robust physical picture and point at the same conclusion.

The main reason for the SF can be understood from the basic electronic structure. The metallicity of the surface and the subsurface CMO layers is due the presence of the $d_{z^2}$ states at the Fermi surface. 
These orbitals point toward each other, which facilitates electron hopping between the Mn ions through a double-exchange-like mechanism and provides a FM coupling. 
This result, in the limit of LSDA, has been also observed in several prior studies~\cite{pickett-2,nguyen}.

An interesting observation in Fig.~\ref{fig:slab-j-unrel} is that LSDA+$U$ predicts larger values for the exchange parameters (red dashed line) which is in contrast with our conclusion for the bulk, where LSDA+$U$ suppressed the LSDA exchange. The reason is due to the metallicity~\cite{super-double} that has been observed here for the surface layers even within LSDA+$U$. In general, the exchange interactions in magnanites can be regarded as the combination of ferromagnetic double exchange and antiferromagnetic superexchange interactions. The superexchange interactions are inversely proportional to U and, therefore, weaker in LSDA+$U$. The double exchange is, on the other hand, proportional to the transfer integral and (to the first approximation) does not depend on U. However, in LSDA+$U$ there are other factors (e.g. the oxygen states), which can alter this conclusion.

The metallic character drastically changes the behavior of the magnetic interactions and the argument based solely on the analysis of the superexchange strength is not anymore valid.

While studying Mn-based perovsikes, one should notice that according to the Bethe-Slater curve~\cite{bethe}, Mn atoms can present drastically modified magnetic properties if the interatomic distances are changed, especially for the confined system in lower dimensions like surfaces. 
This stimulated us to study how the magnetic properties vary once the reconstruction of the surface geometry is taken into account in \textit{ab initio} simulations. 

\subsubsection{The effect of geometry relaxation} 

\begin{figure}[tp]   
 \includegraphics[width=0.36\textwidth]{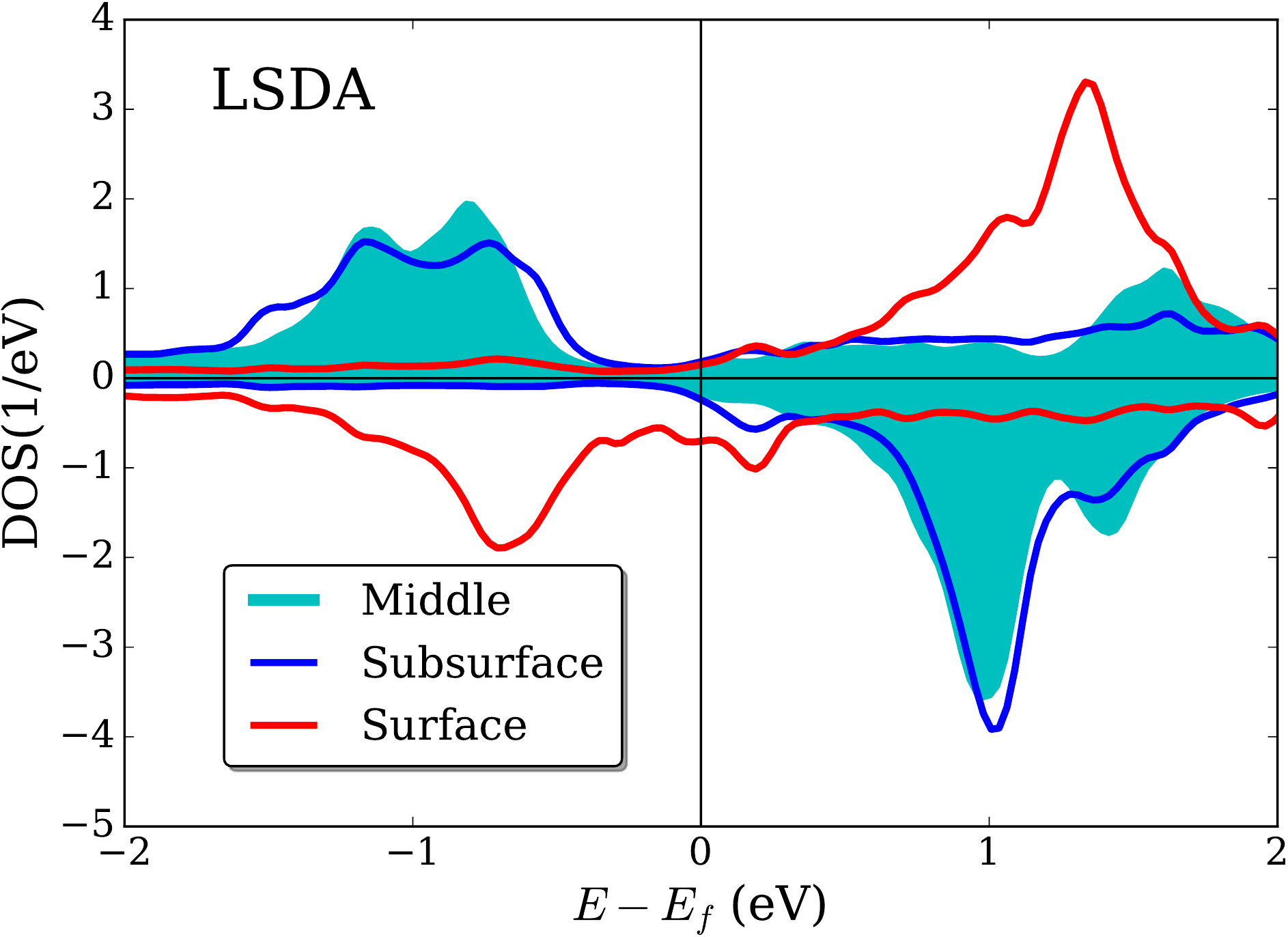}
 \includegraphics[width=0.36\textwidth]{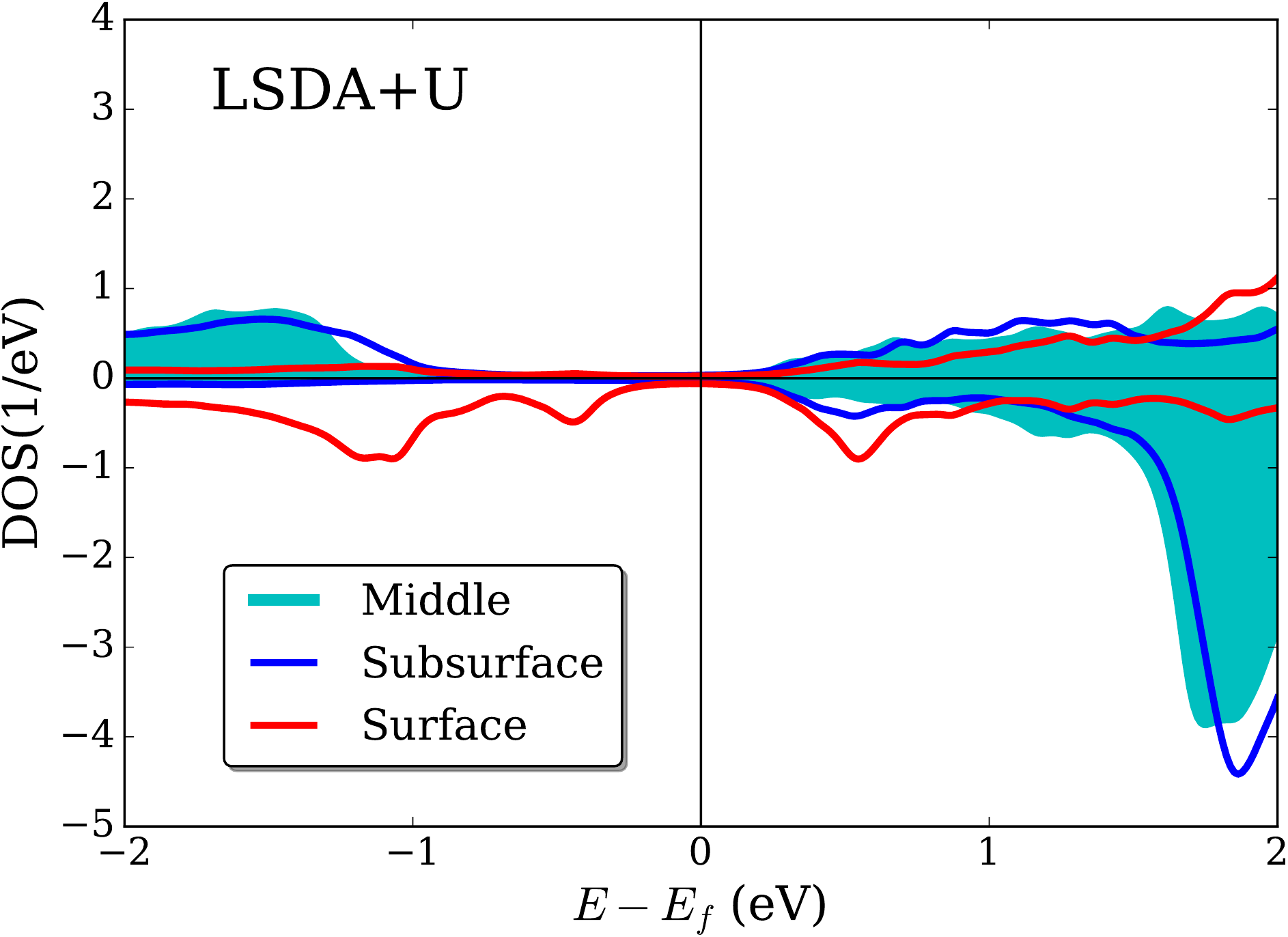}
 \caption{The projected density of states of the relaxed structure of CMO slab for the majority- and minority-spin components in LSDA (top panel) and LSDA+$U$ (bottom panel). G-type AFM ground state.}
\label{fig:slab-dos-rel}
%
 \includegraphics[width=0.4\textwidth]{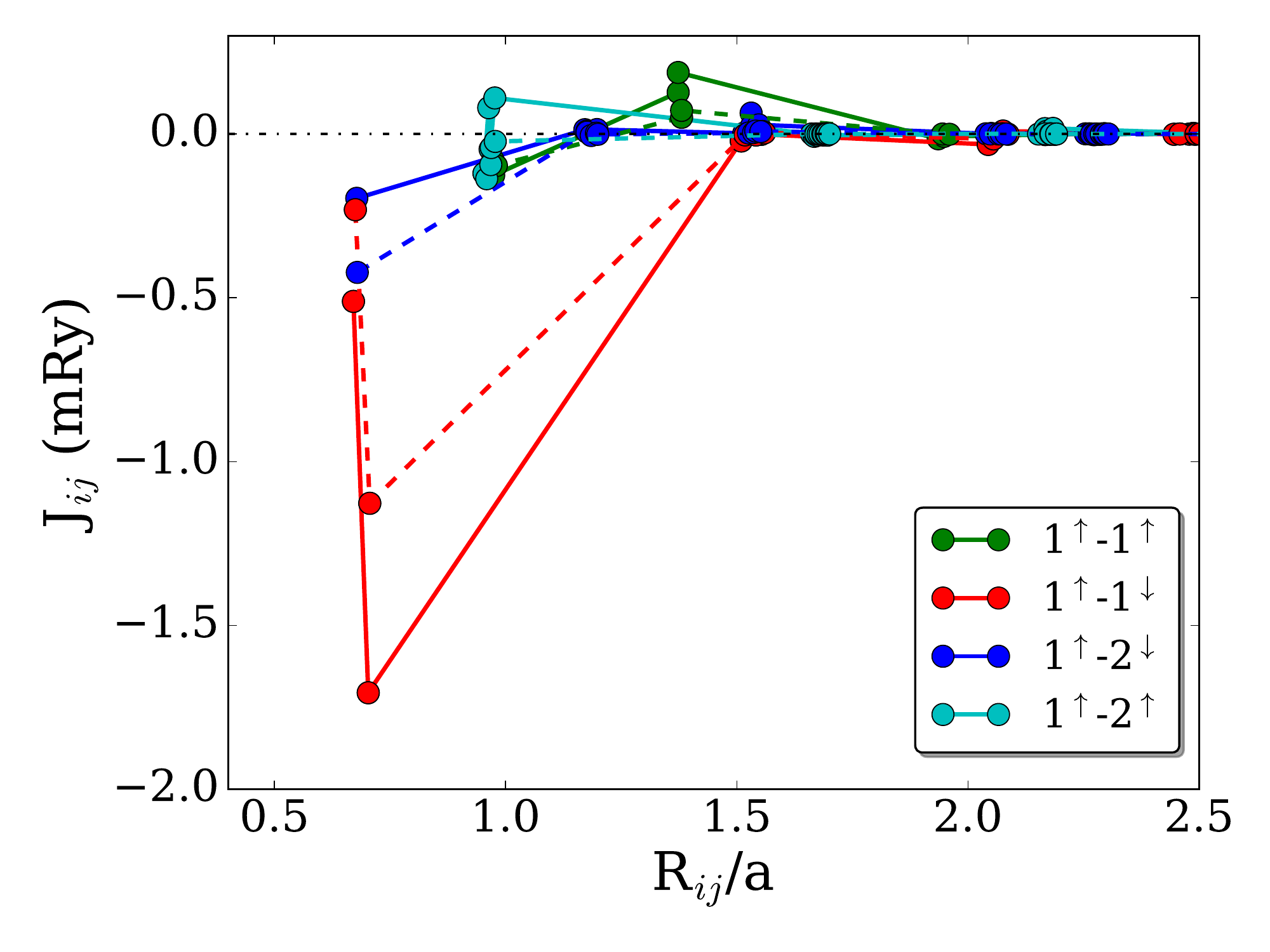}
 \includegraphics[width=0.21\textwidth]{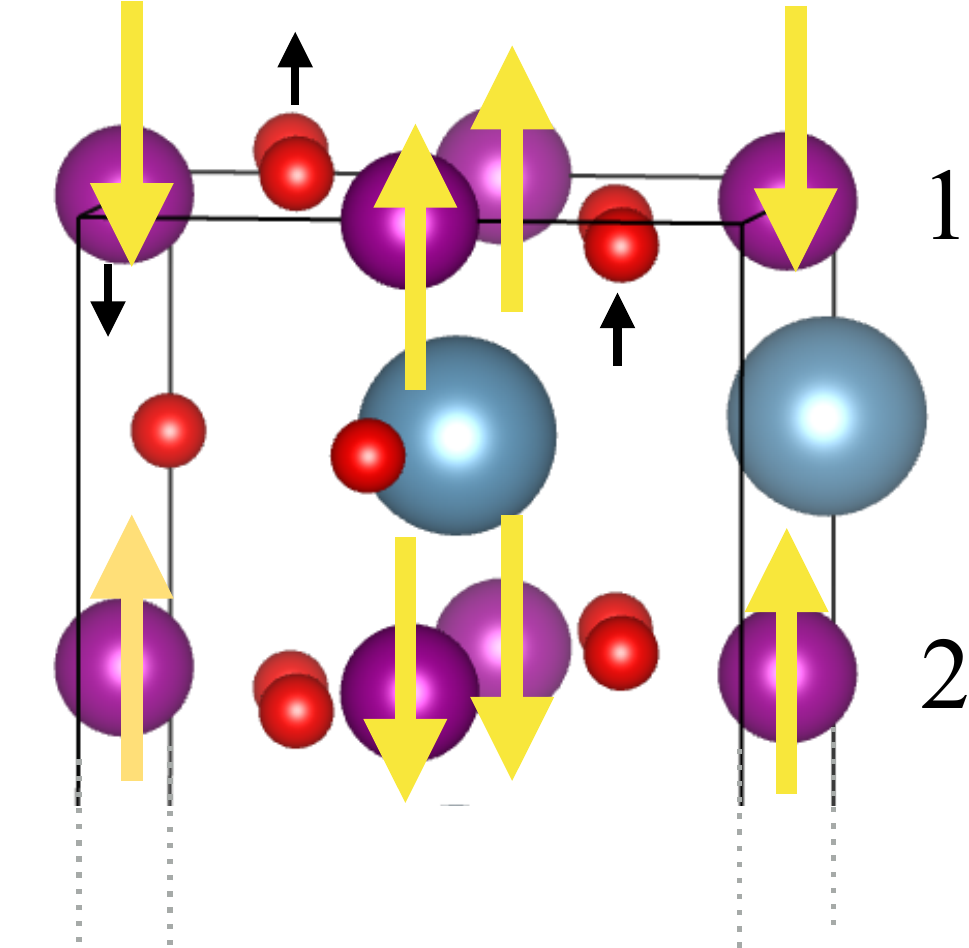}
 \caption{Top panel: The exchange interaction between an atom at the surface of CMO for a relaxed slab with the atoms at the surface (layer 1) and subsurface (layer 2) in LSDA (solid lines) and LSDA+$U$ (dashed lines). Bottom panel: The obtained magnetic order for the relaxed structure. The purple balls represents Mn atoms with yellow arrow indicating spin direction. The black arrows indicate the direction of the movement after geometry relaxation.}
\label{fig:slab-j-rel}
\end{figure}

\begin{figure}[t]   
 \includegraphics[width=0.35\textwidth]{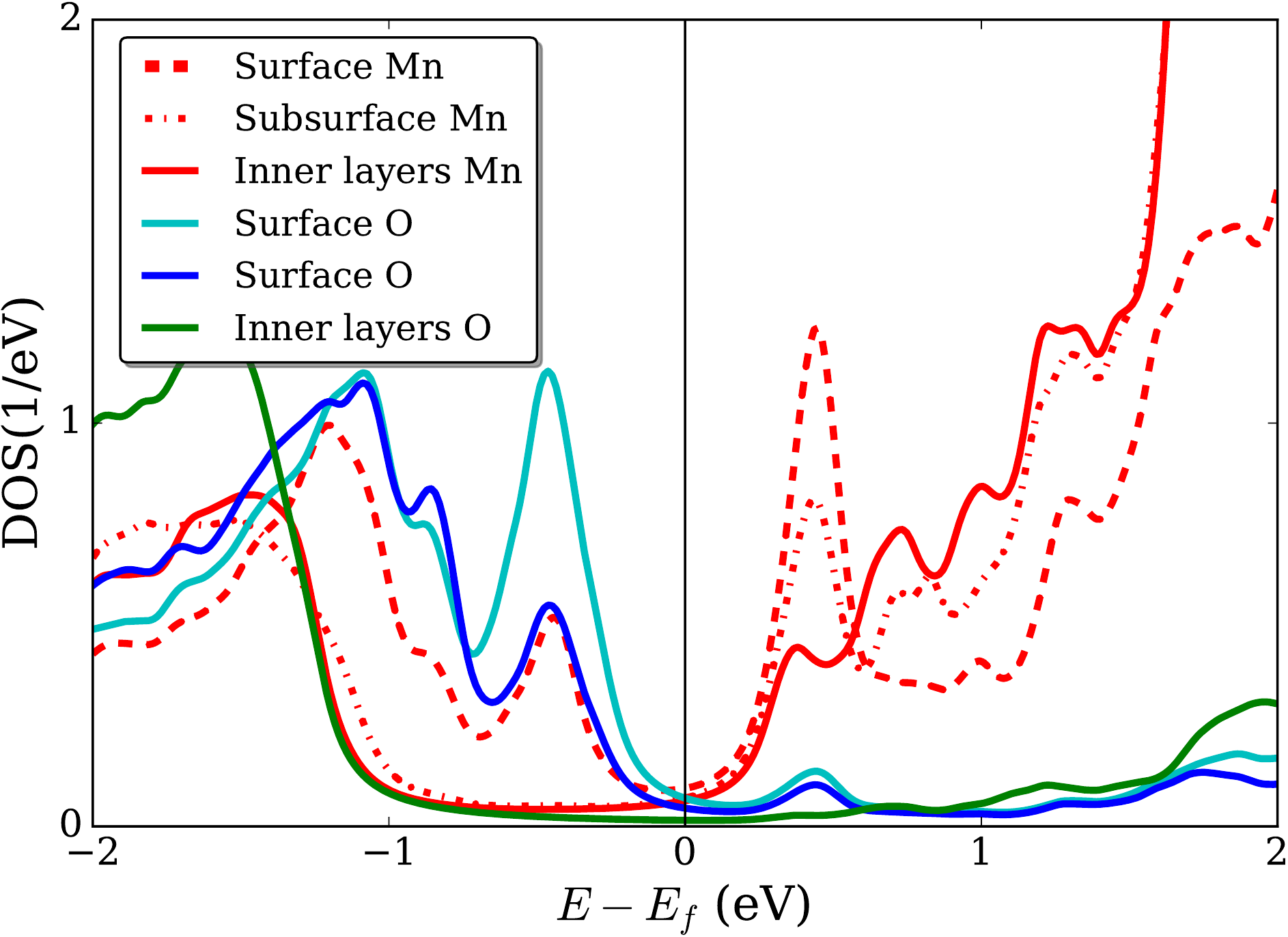}
 \caption{Projected density of states for Mn 3\textit{d} and oxygen 2\textit{p} orbitals calculated in SF configuration after geometrical relaxation within LSDA+$U$.}
\label{fig:slab-sf-rel}
\end{figure}

Next, we performed geometry relaxation for two different spin-configurations of the slab, e.g., AFM and SF at the surface. 
First, we discuss the results obtained for AFM, but the results are qualitatively the same for the other magnetic state. e.g., with SF at the surface.
Once the structure was optimized, we observed a different electronic structure and magnetic properties of the CMO slab compared to the unrelaxed slab. 
Figure~\ref{fig:slab-dos-rel} illustrates the projected density of states for the most physically interesting layers, e.g. surface, subsurface and middle layers in the limit of LSDA and LSDA+$U$. While LSDA gives a metallic ground state, LSDA+$U$ predicts the emergence of a splitting between the Mn $d_{z^2}$ states and O 2\textit{p} states that can turn the system to a charge-transfer insulator. In fact, the lack of periodicity at the surface results in a different crystal field splitting (than for bulk) for Mn \textit{d} states by lifting their degeneracy. In the case of LSDA+$U$, we observed that the electronic states are pushed closer to the Fermi level, while still keeping a clear band gap of 0.4 eV. 
Turning to an insulator, due to the strong structural relaxation and to the change of the crystal field, the system is anticipated to show quite different behavior in the magnetic properties, as has been observed also in other TM perovskites~\cite{carmine}. This change in the electronic structure was found to be independent of the magnetic ground state of the system. For comparison, in Fig.~\ref{fig:slab-sf-rel} we show the calculated DOS for SF configuration at the surface. This also exhibits the opening of the gap.
However, the total energy calculations suggests that once the structural relaxation is taken into account, the preferred magnetic ground state of CMO slab is AFM, different from the unrelaxed geometry with SF at the surface.

Therefore, we repeated the calculations of the exchange parameters starting from the optimized structure of the slabs. The initial magnetic configuration is set to be G-type AFM without SF at the surface. The results for the $J_{ij}$ calculations are illustrated in Fig.~\ref{fig:slab-j-rel}. The exchange parameters also confirm the conclusion that the SF of the surface atoms disappears after the geometry relaxation of the slab (see the blue lines). This conclusion is obtained both in LSDA and in LSDA+$U$ methods. Surprisingly, additional calculations based on G-type AFM in the presence of the SF at the surface also confirms this finding (not shown here).

An inspection of Fig.~\ref{fig:slab-j-rel} reveals that the exchange interaction between the two nearest neighbor Mn atoms at the surface is AFM.
The LSDA approach predicts this interaction to be strong (full red line), whereas it is relatively suppressed in the case of LSDA+$U$ (dashed red line). 
As explained above, this is because the system is metallic in LSDA and other mechanism in addition to superexchange will contribute to the $J_{ij}$'s. 
On the other hand, LSDA+$U$ predicts insulating behaviour and therefore, these extra contributions are suppressed.

Considering that superexchange mechanism has the biggest contribution to the exchange parameters in CMO, one can realize that the change of the nearest neighbor $J_{ij}$ is mainly due to the change of the angle and the distance between Mn and O atoms, induced by the geometry relaxation. The surface and subsurface atoms moved toward the inner layers and in total the atomic distances are decreased compared to the unrelaxed slab, both within LSDA and LSDA+$U$ (and also within some additional simulations based on GGA). Moreover, the angle between Mn and O atoms have been modified depending on the layer they are in. This change is especially large for the O atoms on the surface, where the ones right above the surface are shifted toward the vacuum after relaxation (see the bottom panel of Fig.~\ref{fig:slab-j-rel}) and therefore, the $\widehat{\mbox{Mn-O-Mn}}$ angle is decreased from 156$^{\circ}$ for the unrelaxed structure to 153.1$^{\circ}$ for the relaxed structure.  The Mn-O distance is also decreased from 1.90\AA~ to 1.83\AA. On the other hand, the O atoms right below the surface are moved toward the surface after the relaxation and have modified the $\widehat{\mbox{Mn-O-Mn}}$ angle to 166.1$^{\circ}$ and the Mn-O distance to 1.87\AA. These changes for the inner layers are relatively small so that the innermost layer finds an environment similar to the bulk. Moreover, due to the movement of Mn atoms at the surface toward the subsurface, the angle between Mn atoms at the surface and subsurface and the O between them, is modified from 158$^{\circ}$ to 155$^{\circ}$. Consequently, the exchange interaction with the surface atoms are modified substantially in terms of sign and magnitudes.

Finally, the robustness of our findings is further confirmed by the $J_{ij}$'s extracted from the SF ground state.
The obtained exchange parameters (not shown here) are qualitatively very similar to the ones shown in Fig.~\ref{fig:slab-j-rel}, but with minor differences in their magnitudes. 
For instance, the NN exchange parameter between the atom at the surface and the subsurface is about -0.35 mRy while the corresponding parameter in Fig.~\ref{fig:slab-j-rel} is equal to -0.42 mRy in LSDA+$U$ method. 
The latter indicates an instability of SF ground state and points towards G-type AFM ordering. 

In calculations of the exchange parameters, the SO coupling is not taken into account. However, our additional electronic structure calculations including spin-orbit interactions revealed that the orbital moments originating from the Mn atoms are very small (of the order of 0.01$\mu_B$) both in the bulk and the slab. The canting angle of the spin moments with respect to the spin axis is obtained to be about 0.31$^{\circ}$ for the surface atoms and 0.22$^{\circ}$ for the middle layer atoms in LSDA+$U$ calculations. 
This canting of the spins is a result of purely SO-derived magnetic interactions.
Their smallness allows us to conclude that the influence of SO interaction is marginal, and that the exchange parameters obtained in this work can present the general magnetic behavior of system either in the presence or absence of the spin-orbit couplings. 

\subsubsection{Electron doping}

Excess of electrons can appear in CMO due to doping or they can emerge spontaneously due to the presence of oxygen vacancies.
This can influence the magnetic properties, and therefore we have examined how the electron doping affects the $J_{ij}$'s at the surface. 
For this purpose we have performed an additional set of calculations varying the chemical potential of the system.
Since the converged DFT potential has not been modified, one can best describe this procedure as that the doping was simulated within a rigid band model. 
Note that the present approach is extremely simplified and assumes uniform distribution of an additional negative charge in the system. 
Most importantly, it neglects structural changes leading to the electron trapping, i.e. the formation of magnetic polarons~\cite{cmo-polarons}.
Nonetheless, the aim of these simulations is to provide a qualitative picture about the sensitivity of the obtained preferable magnetic order to the stoichiometry of the system.

\begin{figure}[tp]   
 \includegraphics[width=0.45\textwidth]{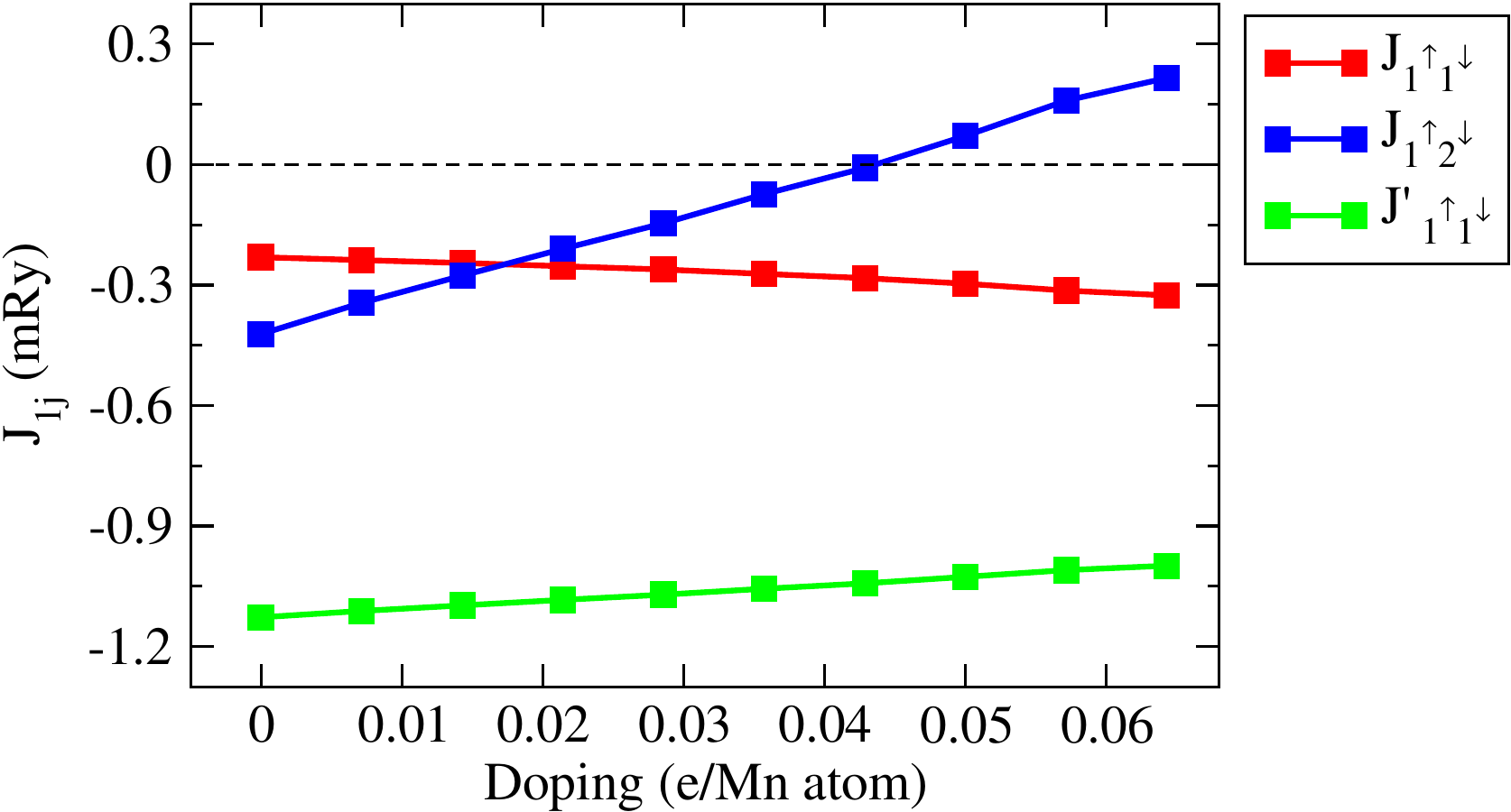}
 \caption{The nearest neighbor exchange parameter as a function of electron doping in the relaxed structure of CMO slab. $J_{1\uparrow 1\downarrow}$ and $J'_{1\uparrow 1\downarrow}$ indicate the exchange interactions between two atoms at the surface and $J_{1\uparrow 2\downarrow}$ is the exchange between the atom at the surface and the one at the subsurface.}
\label{fig:doping}
\end{figure}

Figure~\ref{fig:doping} shows the dependence of the most relevant magnetic couplings at the surface of CMO on electron doping. The nearest neighbor exchange interaction between two atoms at the surface are indicated by two different arguments $J_{1\uparrow 1\downarrow}$ and $J'_{1\uparrow 1\downarrow}$. As mentioned before, this is because of the orthorhombic structure of CMO which makes the x and y axis inequivalent and therefore the distance between the Mn atoms in one direction is 0.17~\AA~ shorter than the other direction. The interaction between the atom at the surface and the one right below in the subsurface is indicated by $J_{1\uparrow 2\downarrow}$, as also depicted in the bottom panel of Fig.~\ref{fig:slab-j-rel}. 
One can see that the interactions between Mn atoms at the surface are relatively robust with respect to the shift of the chemical potential.
In the entire range of doping levels considered here, these couplings remain AFM.
On the other hand, surface-subsurface interaction ($J_{1\uparrow 2\downarrow}$), being AFM for low doping, changes its sign when more that 0.04 electrons per Mn atom are added to the system.
At this point, the SF of Mn spins is expected to take place at the surface of CMO and consequently double exchange mechanism steps in and overrides the super exchange.

\section{Conclusions}
\label{conclusions}

In this paper we have evaluated the exchange interactions of bulk and surface layers of CMO, and we find that super-exchange is captured well in LSDA+$U$ calculations. This conclusion is based on the fact that theoretical exchange interactions combined with an effective spin-Hamiltonian result in a Ne\'el temperature that is in agreement with observations. In the bulk we find that the $t_{2g}-t_{2g}$ channel is dominating. For the surface we find that structural relaxations play an important role, and that when taken into account the (001)-surface of CMO has the same antiferromagnetic ordering as the bulk. This indicates a large exchange-striction in which small structural displacements provide a significant modification of the interatomic exchange interaction, even to the extant that the sign changes. Finally, we predict that a small amount of electron doping of 0.04 electrons per Mn atom, results in a SF transition of the Mn surface atoms. It would be interesting to verify this prediction in experiments and we hope that such investigation will be carried out in the future.

\section*{Acknowledgement}
The authors are grateful to Stefan Bl\"ugel (Forschungszentrum J\"ulich) for stimulating discussions. O.~E. acknowledges support from the Swedish Research Council (VR), eSSENCE and the KAW foundation. The work of IVS is partly supported by the grant of Russian Science Foundation (project No. 14-12-00306). The computer simulations are performed on computational resources provided by NSC and UPPMAX allocated by the Swedish National Infrastructure for Computing (SNIC). D.C.M.R. acknowledges CAPES (Brazil).

\bibliographystyle{apsrev4-1}

\begin{thebibliography}{48}%
\makeatletter
\providecommand \@ifxundefined [1]{%
 \@ifx{#1\undefined}
}%
\providecommand \@ifnum [1]{%
 \ifnum #1\expandafter \@firstoftwo
 \else \expandafter \@secondoftwo
 \fi
}%
\providecommand \@ifx [1]{%
 \ifx #1\expandafter \@firstoftwo
 \else \expandafter \@secondoftwo
 \fi
}%
\providecommand \natexlab [1]{#1}%
\providecommand \enquote  [1]{``#1''}%
\providecommand \bibnamefont  [1]{#1}%
\providecommand \bibfnamefont [1]{#1}%
\providecommand \citenamefont [1]{#1}%
\providecommand \href@noop [0]{\@secondoftwo}%
\providecommand \href [0]{\begingroup \@sanitize@url \@href}%
\providecommand \@href[1]{\@@startlink{#1}\@@href}%
\providecommand \@@href[1]{\endgroup#1\@@endlink}%
\providecommand \@sanitize@url [0]{\catcode `\\12\catcode `\$12\catcode
  `\&12\catcode `\#12\catcode `\^12\catcode `\_12\catcode `\%12\relax}%
\providecommand \@@startlink[1]{}%
\providecommand \@@endlink[0]{}%
\providecommand \url  [0]{\begingroup\@sanitize@url \@url }%
\providecommand \@url [1]{\endgroup\@href {#1}{\urlprefix }}%
\providecommand \urlprefix  [0]{URL }%
\providecommand \Eprint [0]{\href }%
\providecommand \doibase [0]{http://dx.doi.org/}%
\providecommand \selectlanguage [0]{\@gobble}%
\providecommand \bibinfo  [0]{\@secondoftwo}%
\providecommand \bibfield  [0]{\@secondoftwo}%
\providecommand \translation [1]{[#1]}%
\providecommand \BibitemOpen [0]{}%
\providecommand \bibitemStop [0]{}%
\providecommand \bibitemNoStop [0]{.\EOS\space}%
\providecommand \EOS [0]{\spacefactor3000\relax}%
\providecommand \BibitemShut  [1]{\csname bibitem#1\endcsname}%
\let\auto@bib@innerbib\@empty
\bibitem [{\citenamefont {Dagotto}(2013)}]{dagotto-book}%
  \BibitemOpen
  \bibfield  {author} {\bibinfo {author} {\bibfnamefont {E.}~\bibnamefont
  {Dagotto}},\ }\href {https://books.google.se/books?id=rS7qCAAAQBAJ} {\emph
  {\bibinfo {title} {Nanoscale Phase Separation and Colossal Magnetoresistance:
  The Physics of Manganites and Related Compounds}}},\ Springer Series in
  Solid-State Sciences\ (\bibinfo  {publisher} {Springer Berlin Heidelberg},\
  \bibinfo {year} {2013})\BibitemShut {NoStop}%
\bibitem [{\citenamefont {Sanyal}\ and\ \citenamefont
  {Eriksson}(2012)}]{BS-and-OE-book}%
  \BibitemOpen
  \bibfield  {author} {\bibinfo {author} {\bibfnamefont {B.}~\bibnamefont
  {Sanyal}}\ and\ \bibinfo {author} {\bibfnamefont {O.}~\bibnamefont
  {Eriksson}},\ }\href {https://books.google.se/books?id=r99rAQkxLJoC} {\emph
  {\bibinfo {title} {Advanced Functional Materials: A Perspective from Theory
  and Experiment}}},\ Science and Technology of Atomic, Molecular, Condensed
  Matter \& Biological Systems\ (\bibinfo  {publisher} {Elsevier Science},\
  \bibinfo {year} {2012})\BibitemShut {NoStop}%
\bibitem [{\citenamefont {Bhattacharjee}\ \emph {et~al.}(2009)\citenamefont
  {Bhattacharjee}, \citenamefont {Bousquet},\ and\ \citenamefont
  {Ghosez}}]{eric}%
  \BibitemOpen
  \bibfield  {author} {\bibinfo {author} {\bibfnamefont {S.}~\bibnamefont
  {Bhattacharjee}}, \bibinfo {author} {\bibfnamefont {E.}~\bibnamefont
  {Bousquet}}, \ and\ \bibinfo {author} {\bibfnamefont {P.}~\bibnamefont
  {Ghosez}},\ }\href {\doibase 10.1103/PhysRevLett.102.117602} {\bibfield
  {journal} {\bibinfo  {journal} {Phys. Rev. Lett.}\ }\textbf {\bibinfo
  {volume} {102}},\ \bibinfo {pages} {117602} (\bibinfo {year}
  {2009})}\BibitemShut {NoStop}%
\bibitem [{\citenamefont {Cheong}\ and\ \citenamefont
  {Mostovoy}(2007)}]{TbMnO3-Mostovoy}%
  \BibitemOpen
  \bibfield  {author} {\bibinfo {author} {\bibfnamefont {S.-W.}\ \bibnamefont
  {Cheong}}\ and\ \bibinfo {author} {\bibfnamefont {M.}~\bibnamefont
  {Mostovoy}},\ }\href {\doibase 10.1038/nmat1804} {\bibfield  {journal}
  {\bibinfo  {journal} {Nature Materials}\ }\textbf {\bibinfo {volume} {6}},\
  \bibinfo {pages} {13} (\bibinfo {year} {2007})}\BibitemShut {NoStop}%
\bibitem [{\citenamefont {Schiffer}\ \emph {et~al.}(1995)\citenamefont
  {Schiffer}, \citenamefont {Ramirez}, \citenamefont {Bao},\ and\ \citenamefont
  {Cheong}}]{CMR-LCMO}%
  \BibitemOpen
  \bibfield  {author} {\bibinfo {author} {\bibfnamefont {P.}~\bibnamefont
  {Schiffer}}, \bibinfo {author} {\bibfnamefont {A.~P.}\ \bibnamefont
  {Ramirez}}, \bibinfo {author} {\bibfnamefont {W.}~\bibnamefont {Bao}}, \ and\
  \bibinfo {author} {\bibfnamefont {S.-W.}\ \bibnamefont {Cheong}},\ }\href
  {\doibase 10.1103/PhysRevLett.75.3336} {\bibfield  {journal} {\bibinfo
  {journal} {Phys. Rev. Lett.}\ }\textbf {\bibinfo {volume} {75}},\ \bibinfo
  {pages} {3336} (\bibinfo {year} {1995})}\BibitemShut {NoStop}%
\bibitem [{\citenamefont {Reyren}\ \emph {et~al.}(2007)\citenamefont {Reyren},
  \citenamefont {Thiel}, \citenamefont {Caviglia}, \citenamefont {Kourkoutis},
  \citenamefont {Hammerl}, \citenamefont {Richter}, \citenamefont {Schneider},
  \citenamefont {Kopp}, \citenamefont {R{\"u}etschi}, \citenamefont {Jaccard},
  \citenamefont {Gabay}, \citenamefont {Muller}, \citenamefont {Triscone},\
  and\ \citenamefont {Mannhart}}]{lao-sto-supercond}%
  \BibitemOpen
  \bibfield  {author} {\bibinfo {author} {\bibfnamefont {N.}~\bibnamefont
  {Reyren}}, \bibinfo {author} {\bibfnamefont {S.}~\bibnamefont {Thiel}},
  \bibinfo {author} {\bibfnamefont {A.~D.}\ \bibnamefont {Caviglia}}, \bibinfo
  {author} {\bibfnamefont {L.~F.}\ \bibnamefont {Kourkoutis}}, \bibinfo
  {author} {\bibfnamefont {G.}~\bibnamefont {Hammerl}}, \bibinfo {author}
  {\bibfnamefont {C.}~\bibnamefont {Richter}}, \bibinfo {author} {\bibfnamefont
  {C.~W.}\ \bibnamefont {Schneider}}, \bibinfo {author} {\bibfnamefont
  {T.}~\bibnamefont {Kopp}}, \bibinfo {author} {\bibfnamefont {A.-S.}\
  \bibnamefont {R{\"u}etschi}}, \bibinfo {author} {\bibfnamefont
  {D.}~\bibnamefont {Jaccard}}, \bibinfo {author} {\bibfnamefont
  {M.}~\bibnamefont {Gabay}}, \bibinfo {author} {\bibfnamefont {D.~A.}\
  \bibnamefont {Muller}}, \bibinfo {author} {\bibfnamefont {J.-M.}\
  \bibnamefont {Triscone}}, \ and\ \bibinfo {author} {\bibfnamefont
  {J.}~\bibnamefont {Mannhart}},\ }\href {\doibase 10.1126/science.1146006} {\
  \textbf {\bibinfo {volume} {317}},\ \bibinfo {pages} {1196} (\bibinfo {year}
  {2007})}\BibitemShut {NoStop}%
\bibitem [{\citenamefont {Ohtomo}\ and\ \citenamefont {Hwang}(2004)}]{2deg}%
  \BibitemOpen
  \bibfield  {author} {\bibinfo {author} {\bibfnamefont {A.}~\bibnamefont
  {Ohtomo}}\ and\ \bibinfo {author} {\bibfnamefont {H.}~\bibnamefont {Hwang}},\
  }\href@noop {} {\bibfield  {journal} {\bibinfo  {journal} {Nature}\ }\textbf
  {\bibinfo {volume} {427}},\ \bibinfo {pages} {423} (\bibinfo {year}
  {2004})}\BibitemShut {NoStop}%
\bibitem [{\citenamefont {Meevasana}\ \emph {et~al.}(2011)\citenamefont
  {Meevasana}, \citenamefont {King}, \citenamefont {He}, \citenamefont {Mo},
  \citenamefont {Hashimoto}, \citenamefont {Tamai}, \citenamefont
  {Songsiriritthigul}, \citenamefont {Baumberger},\ and\ \citenamefont
  {Shen}}]{surf-2deg}%
  \BibitemOpen
  \bibfield  {author} {\bibinfo {author} {\bibfnamefont {W.}~\bibnamefont
  {Meevasana}}, \bibinfo {author} {\bibfnamefont {P.~D.~C.}\ \bibnamefont
  {King}}, \bibinfo {author} {\bibfnamefont {R.~H.}\ \bibnamefont {He}},
  \bibinfo {author} {\bibfnamefont {S.-K.}\ \bibnamefont {Mo}}, \bibinfo
  {author} {\bibfnamefont {M.}~\bibnamefont {Hashimoto}}, \bibinfo {author}
  {\bibfnamefont {A.}~\bibnamefont {Tamai}}, \bibinfo {author} {\bibfnamefont
  {P.}~\bibnamefont {Songsiriritthigul}}, \bibinfo {author} {\bibfnamefont
  {F.}~\bibnamefont {Baumberger}}, \ and\ \bibinfo {author} {\bibfnamefont
  {Z.-X.}\ \bibnamefont {Shen}},\ }\href {http://dx.doi.org/10.1038/nmat2943}
  {\bibfield  {journal} {\bibinfo  {journal} {Nat Mater}\ }\textbf {\bibinfo
  {volume} {10}},\ \bibinfo {pages} {114} (\bibinfo {year} {2011})}\BibitemShut
  {NoStop}%
\bibitem [{\citenamefont {Ramirez}\ \emph {et~al.}(1996)\citenamefont
  {Ramirez}, \citenamefont {Schiffer}, \citenamefont {Cheong}, \citenamefont
  {Chen}, \citenamefont {Bao}, \citenamefont {Palstra}, \citenamefont {Gammel},
  \citenamefont {Bishop},\ and\ \citenamefont {Zegarski}}]{lcmo-phasediag}%
  \BibitemOpen
  \bibfield  {author} {\bibinfo {author} {\bibfnamefont {A.~P.}\ \bibnamefont
  {Ramirez}}, \bibinfo {author} {\bibfnamefont {P.}~\bibnamefont {Schiffer}},
  \bibinfo {author} {\bibfnamefont {S.-W.}\ \bibnamefont {Cheong}}, \bibinfo
  {author} {\bibfnamefont {C.~H.}\ \bibnamefont {Chen}}, \bibinfo {author}
  {\bibfnamefont {W.}~\bibnamefont {Bao}}, \bibinfo {author} {\bibfnamefont
  {T.~T.~M.}\ \bibnamefont {Palstra}}, \bibinfo {author} {\bibfnamefont
  {P.~L.}\ \bibnamefont {Gammel}}, \bibinfo {author} {\bibfnamefont {D.~J.}\
  \bibnamefont {Bishop}}, \ and\ \bibinfo {author} {\bibfnamefont
  {B.}~\bibnamefont {Zegarski}},\ }\href {\doibase 10.1103/PhysRevLett.76.3188}
  {\bibfield  {journal} {\bibinfo  {journal} {Phys. Rev. Lett.}\ }\textbf
  {\bibinfo {volume} {76}},\ \bibinfo {pages} {3188} (\bibinfo {year}
  {1996})}\BibitemShut {NoStop}%
\bibitem [{\citenamefont {Neumeier}\ and\ \citenamefont
  {Cohn}(2000)}]{cmo-polarons}%
  \BibitemOpen
  \bibfield  {author} {\bibinfo {author} {\bibfnamefont {J.~J.}\ \bibnamefont
  {Neumeier}}\ and\ \bibinfo {author} {\bibfnamefont {J.~L.}\ \bibnamefont
  {Cohn}},\ }\href {\doibase 10.1103/PhysRevB.61.14319} {\bibfield  {journal}
  {\bibinfo  {journal} {Phys. Rev. B}\ }\textbf {\bibinfo {volume} {61}},\
  \bibinfo {pages} {14319} (\bibinfo {year} {2000})}\BibitemShut {NoStop}%
\bibitem [{\citenamefont {Barone}\ \emph {et~al.}(2014)\citenamefont {Barone},
  \citenamefont {Di~Sante},\ and\ \citenamefont {Picozzi}}]{silvia}%
  \BibitemOpen
  \bibfield  {author} {\bibinfo {author} {\bibfnamefont {P.}~\bibnamefont
  {Barone}}, \bibinfo {author} {\bibfnamefont {D.}~\bibnamefont {Di~Sante}}, \
  and\ \bibinfo {author} {\bibfnamefont {S.}~\bibnamefont {Picozzi}},\ }\href
  {\doibase 10.1103/PhysRevB.89.144104} {\bibfield  {journal} {\bibinfo
  {journal} {Phys. Rev. B}\ }\textbf {\bibinfo {volume} {89}},\ \bibinfo
  {pages} {144104} (\bibinfo {year} {2014})}\BibitemShut {NoStop}%
\bibitem [{\citenamefont {Anderson}(1959)}]{pwa-superexch}%
  \BibitemOpen
  \bibfield  {author} {\bibinfo {author} {\bibfnamefont {P.~W.}\ \bibnamefont
  {Anderson}},\ }\href {\doibase 10.1103/PhysRev.115.2} {\bibfield  {journal}
  {\bibinfo  {journal} {Phys. Rev.}\ }\textbf {\bibinfo {volume} {115}},\
  \bibinfo {pages} {2} (\bibinfo {year} {1959})}\BibitemShut {NoStop}%
\bibitem [{\citenamefont {Loshkareva}\ \emph {et~al.}(2004)\citenamefont
  {Loshkareva}, \citenamefont {Nomerovannaya}, \citenamefont {Mostovshchikova},
  \citenamefont {Makhnev}, \citenamefont {Sukhorukov}, \citenamefont {Solin},
  \citenamefont {Arbuzova}, \citenamefont {Naumov}, \citenamefont
  {Kostromitina}, \citenamefont {Balbashov},\ and\ \citenamefont
  {Rybina}}]{optical-gap}%
  \BibitemOpen
  \bibfield  {author} {\bibinfo {author} {\bibfnamefont {N.~N.}\ \bibnamefont
  {Loshkareva}}, \bibinfo {author} {\bibfnamefont {L.~V.}\ \bibnamefont
  {Nomerovannaya}}, \bibinfo {author} {\bibfnamefont {E.~V.}\ \bibnamefont
  {Mostovshchikova}}, \bibinfo {author} {\bibfnamefont {A.~A.}\ \bibnamefont
  {Makhnev}}, \bibinfo {author} {\bibfnamefont {Y.~P.}\ \bibnamefont
  {Sukhorukov}}, \bibinfo {author} {\bibfnamefont {N.~I.}\ \bibnamefont
  {Solin}}, \bibinfo {author} {\bibfnamefont {T.~I.}\ \bibnamefont {Arbuzova}},
  \bibinfo {author} {\bibfnamefont {S.~V.}\ \bibnamefont {Naumov}}, \bibinfo
  {author} {\bibfnamefont {N.~V.}\ \bibnamefont {Kostromitina}}, \bibinfo
  {author} {\bibfnamefont {A.~M.}\ \bibnamefont {Balbashov}}, \ and\ \bibinfo
  {author} {\bibfnamefont {L.~N.}\ \bibnamefont {Rybina}},\ }\href {\doibase
  10.1103/PhysRevB.70.224406} {\bibfield  {journal} {\bibinfo  {journal} {Phys.
  Rev. B}\ }\textbf {\bibinfo {volume} {70}},\ \bibinfo {pages} {224406}
  (\bibinfo {year} {2004})}\BibitemShut {NoStop}%
\bibitem [{\citenamefont {Zeng}\ \emph {et~al.}(1999)\citenamefont {Zeng},
  \citenamefont {Greenblatt},\ and\ \citenamefont {Croft}}]{distortion}%
  \BibitemOpen
  \bibfield  {author} {\bibinfo {author} {\bibfnamefont {Z.}~\bibnamefont
  {Zeng}}, \bibinfo {author} {\bibfnamefont {M.}~\bibnamefont {Greenblatt}}, \
  and\ \bibinfo {author} {\bibfnamefont {M.}~\bibnamefont {Croft}},\ }\href
  {\doibase 10.1103/PhysRevB.59.8784} {\bibfield  {journal} {\bibinfo
  {journal} {Phys. Rev. B}\ }\textbf {\bibinfo {volume} {59}},\ \bibinfo
  {pages} {8784} (\bibinfo {year} {1999})}\BibitemShut {NoStop}%
\bibitem [{\citenamefont {Zener}(1951)}]{zener-de}%
  \BibitemOpen
  \bibfield  {author} {\bibinfo {author} {\bibfnamefont {C.}~\bibnamefont
  {Zener}},\ }\href {\doibase 10.1103/PhysRev.82.403} {\bibfield  {journal}
  {\bibinfo  {journal} {Phys. Rev.}\ }\textbf {\bibinfo {volume} {82}},\
  \bibinfo {pages} {403} (\bibinfo {year} {1951})}\BibitemShut {NoStop}%
\bibitem [{\citenamefont {de~Gennes}(1960)}]{double-1}%
  \BibitemOpen
  \bibfield  {author} {\bibinfo {author} {\bibfnamefont {P.~G.}\ \bibnamefont
  {de~Gennes}},\ }\href {\doibase 10.1103/PhysRev.118.141} {\bibfield
  {journal} {\bibinfo  {journal} {Phys. Rev.}\ }\textbf {\bibinfo {volume}
  {118}},\ \bibinfo {pages} {141} (\bibinfo {year} {1960})}\BibitemShut
  {NoStop}%
\bibitem [{\citenamefont {Anderson}\ and\ \citenamefont
  {Hasegawa}(1955)}]{double-2}%
  \BibitemOpen
  \bibfield  {author} {\bibinfo {author} {\bibfnamefont {P.~W.}\ \bibnamefont
  {Anderson}}\ and\ \bibinfo {author} {\bibfnamefont {H.}~\bibnamefont
  {Hasegawa}},\ }\href {\doibase 10.1103/PhysRev.100.675} {\bibfield  {journal}
  {\bibinfo  {journal} {Phys. Rev.}\ }\textbf {\bibinfo {volume} {100}},\
  \bibinfo {pages} {675} (\bibinfo {year} {1955})}\BibitemShut {NoStop}%
\bibitem [{\citenamefont {Filippetti}\ and\ \citenamefont
  {Pickett}(1999)}]{pickett-1}%
  \BibitemOpen
  \bibfield  {author} {\bibinfo {author} {\bibfnamefont {A.}~\bibnamefont
  {Filippetti}}\ and\ \bibinfo {author} {\bibfnamefont {W.~E.}\ \bibnamefont
  {Pickett}},\ }\href {\doibase 10.1103/PhysRevLett.83.4184} {\bibfield
  {journal} {\bibinfo  {journal} {Phys. Rev. Lett.}\ }\textbf {\bibinfo
  {volume} {83}},\ \bibinfo {pages} {4184} (\bibinfo {year}
  {1999})}\BibitemShut {NoStop}%
\bibitem [{\citenamefont {Anisimov}\ \emph {et~al.}(1997)\citenamefont
  {Anisimov}, \citenamefont {Aryasetiawan},\ and\ \citenamefont
  {Lichtenstein}}]{stoner}%
  \BibitemOpen
  \bibfield  {author} {\bibinfo {author} {\bibfnamefont {V.~I.}\ \bibnamefont
  {Anisimov}}, \bibinfo {author} {\bibfnamefont {F.}~\bibnamefont
  {Aryasetiawan}}, \ and\ \bibinfo {author} {\bibfnamefont {A.~I.}\
  \bibnamefont {Lichtenstein}},\ }\href
  {http://stacks.iop.org/0953-8984/9/i=4/a=002} {\bibfield  {journal} {\bibinfo
   {journal} {Journal of Physics: Condensed Matter}\ }\textbf {\bibinfo
  {volume} {9}},\ \bibinfo {pages} {767} (\bibinfo {year} {1997})}\BibitemShut
  {NoStop}%
\bibitem [{\citenamefont {Kresse}\ and\ \citenamefont {Joubert}(1999)}]{paw}%
  \BibitemOpen
  \bibfield  {author} {\bibinfo {author} {\bibfnamefont {G.}~\bibnamefont
  {Kresse}}\ and\ \bibinfo {author} {\bibfnamefont {D.}~\bibnamefont
  {Joubert}},\ }\href {\doibase 10.1103/PhysRevB.59.1758} {\bibfield  {journal}
  {\bibinfo  {journal} {Phys. Rev. B}\ }\textbf {\bibinfo {volume} {59}},\
  \bibinfo {pages} {1758} (\bibinfo {year} {1999})}\BibitemShut {NoStop}%
\bibitem [{\citenamefont {Kresse}\ and\ \citenamefont
  {Furthm{\"u}ller}(1996)}]{vasp}%
  \BibitemOpen
  \bibfield  {author} {\bibinfo {author} {\bibfnamefont {G.}~\bibnamefont
  {Kresse}}\ and\ \bibinfo {author} {\bibfnamefont {J.}~\bibnamefont
  {Furthm{\"u}ller}},\ }\href {\doibase
  http://dx.doi.org/10.1016/0927-0256(96)00008-0} {\bibfield  {journal}
  {\bibinfo  {journal} {Computational Materials Science}\ }\textbf {\bibinfo
  {volume} {6}},\ \bibinfo {pages} {15 } (\bibinfo {year} {1996})}\BibitemShut
  {NoStop}%
\bibitem [{\citenamefont {Perdew}\ and\ \citenamefont {Wang}(1992)}]{lda}%
  \BibitemOpen
  \bibfield  {author} {\bibinfo {author} {\bibfnamefont {J.~P.}\ \bibnamefont
  {Perdew}}\ and\ \bibinfo {author} {\bibfnamefont {Y.}~\bibnamefont {Wang}},\
  }\href {\doibase 10.1103/PhysRevB.45.13244} {\bibfield  {journal} {\bibinfo
  {journal} {Phys. Rev. B}\ }\textbf {\bibinfo {volume} {45}},\ \bibinfo
  {pages} {13244} (\bibinfo {year} {1992})}\BibitemShut {NoStop}%
\bibitem [{\citenamefont {Filippetti}\ and\ \citenamefont
  {Pickett}(2000)}]{pickett-2}%
  \BibitemOpen
  \bibfield  {author} {\bibinfo {author} {\bibfnamefont {A.}~\bibnamefont
  {Filippetti}}\ and\ \bibinfo {author} {\bibfnamefont {W.~E.}\ \bibnamefont
  {Pickett}},\ }\href {\doibase 10.1103/PhysRevB.62.11571} {\bibfield
  {journal} {\bibinfo  {journal} {Phys. Rev. B}\ }\textbf {\bibinfo {volume}
  {62}},\ \bibinfo {pages} {11571} (\bibinfo {year} {2000})}\BibitemShut
  {NoStop}%
\bibitem [{\citenamefont {Hong}\ \emph {et~al.}(2012)\citenamefont {Hong},
  \citenamefont {Stroppa}, \citenamefont {\'I\~niguez}, \citenamefont
  {Picozzi},\ and\ \citenamefont {Vanderbilt}}]{u-j}%
  \BibitemOpen
  \bibfield  {author} {\bibinfo {author} {\bibfnamefont {J.}~\bibnamefont
  {Hong}}, \bibinfo {author} {\bibfnamefont {A.}~\bibnamefont {Stroppa}},
  \bibinfo {author} {\bibfnamefont {J.}~\bibnamefont {\'I\~niguez}}, \bibinfo
  {author} {\bibfnamefont {S.}~\bibnamefont {Picozzi}}, \ and\ \bibinfo
  {author} {\bibfnamefont {D.}~\bibnamefont {Vanderbilt}},\ }\href {\doibase
  10.1103/PhysRevB.85.054417} {\bibfield  {journal} {\bibinfo  {journal} {Phys.
  Rev. B}\ }\textbf {\bibinfo {volume} {85}},\ \bibinfo {pages} {054417}
  (\bibinfo {year} {2012})}\BibitemShut {NoStop}%
\bibitem [{\citenamefont {Jung}\ \emph {et~al.}(1997)\citenamefont {Jung},
  \citenamefont {Kim}, \citenamefont {Eom}, \citenamefont {Noh}, \citenamefont
  {Choi}, \citenamefont {Yu}, \citenamefont {Kwon},\ and\ \citenamefont
  {Chung}}]{optic}%
  \BibitemOpen
  \bibfield  {author} {\bibinfo {author} {\bibfnamefont {J.~H.}\ \bibnamefont
  {Jung}}, \bibinfo {author} {\bibfnamefont {K.~H.}\ \bibnamefont {Kim}},
  \bibinfo {author} {\bibfnamefont {D.~J.}\ \bibnamefont {Eom}}, \bibinfo
  {author} {\bibfnamefont {T.~W.}\ \bibnamefont {Noh}}, \bibinfo {author}
  {\bibfnamefont {E.~J.}\ \bibnamefont {Choi}}, \bibinfo {author}
  {\bibfnamefont {J.}~\bibnamefont {Yu}}, \bibinfo {author} {\bibfnamefont
  {Y.~S.}\ \bibnamefont {Kwon}}, \ and\ \bibinfo {author} {\bibfnamefont
  {Y.}~\bibnamefont {Chung}},\ }\href {\doibase 10.1103/PhysRevB.55.15489}
  {\bibfield  {journal} {\bibinfo  {journal} {Phys. Rev. B}\ }\textbf {\bibinfo
  {volume} {55}},\ \bibinfo {pages} {15489} (\bibinfo {year}
  {1997})}\BibitemShut {NoStop}%
\bibitem [{\citenamefont {Aschauer}\ \emph {et~al.}(2013)\citenamefont
  {Aschauer}, \citenamefont {Pfenninger}, \citenamefont {Selbach},
  \citenamefont {Grande},\ and\ \citenamefont {Spaldin}}]{gap-1}%
  \BibitemOpen
  \bibfield  {author} {\bibinfo {author} {\bibfnamefont {U.}~\bibnamefont
  {Aschauer}}, \bibinfo {author} {\bibfnamefont {R.}~\bibnamefont
  {Pfenninger}}, \bibinfo {author} {\bibfnamefont {S.~M.}\ \bibnamefont
  {Selbach}}, \bibinfo {author} {\bibfnamefont {T.}~\bibnamefont {Grande}}, \
  and\ \bibinfo {author} {\bibfnamefont {N.~A.}\ \bibnamefont {Spaldin}},\
  }\href {\doibase 10.1103/PhysRevB.88.054111} {\bibfield  {journal} {\bibinfo
  {journal} {Phys. Rev. B}\ }\textbf {\bibinfo {volume} {88}},\ \bibinfo
  {pages} {054111} (\bibinfo {year} {2013})}\BibitemShut {NoStop}%
\bibitem [{\citenamefont {Liechtenstein}\ \emph {et~al.}(1995)\citenamefont
  {Liechtenstein}, \citenamefont {Anisimov},\ and\ \citenamefont
  {Zaanen}}]{FLL-DC}%
  \BibitemOpen
  \bibfield  {author} {\bibinfo {author} {\bibfnamefont {A.~I.}\ \bibnamefont
  {Liechtenstein}}, \bibinfo {author} {\bibfnamefont {V.~I.}\ \bibnamefont
  {Anisimov}}, \ and\ \bibinfo {author} {\bibfnamefont {J.}~\bibnamefont
  {Zaanen}},\ }\href {\doibase 10.1103/PhysRevB.52.R5467} {\bibfield  {journal}
  {\bibinfo  {journal} {Phys. Rev. B}\ }\textbf {\bibinfo {volume} {52}},\
  \bibinfo {pages} {R5467} (\bibinfo {year} {1995})}\BibitemShut {NoStop}%
\bibitem [{\citenamefont {Solovyev}\ \emph {et~al.}(1994)\citenamefont
  {Solovyev}, \citenamefont {Dederichs},\ and\ \citenamefont
  {Anisimov}}]{amf-2}%
  \BibitemOpen
  \bibfield  {author} {\bibinfo {author} {\bibfnamefont {I.~V.}\ \bibnamefont
  {Solovyev}}, \bibinfo {author} {\bibfnamefont {P.~H.}\ \bibnamefont
  {Dederichs}}, \ and\ \bibinfo {author} {\bibfnamefont {V.~I.}\ \bibnamefont
  {Anisimov}},\ }\href {\doibase 10.1103/PhysRevB.50.16861} {\bibfield
  {journal} {\bibinfo  {journal} {Phys. Rev. B}\ }\textbf {\bibinfo {volume}
  {50}},\ \bibinfo {pages} {16861} (\bibinfo {year} {1994})}\BibitemShut
  {NoStop}%
\bibitem [{\citenamefont {Wills}\ \emph {et~al.}(2010)\citenamefont {Wills},
  \citenamefont {Alouani}, \citenamefont {Andersson}, \citenamefont {Delin},
  \citenamefont {Eriksson},\ and\ \citenamefont {O.}}]{rspt}%
  \BibitemOpen
  \bibfield  {author} {\bibinfo {author} {\bibfnamefont {J.~M.}\ \bibnamefont
  {Wills}}, \bibinfo {author} {\bibfnamefont {M.}~\bibnamefont {Alouani}},
  \bibinfo {author} {\bibfnamefont {P.}~\bibnamefont {Andersson}}, \bibinfo
  {author} {\bibfnamefont {A.}~\bibnamefont {Delin}}, \bibinfo {author}
  {\bibfnamefont {O.}~\bibnamefont {Eriksson}}, \ and\ \bibinfo {author}
  {\bibfnamefont {G.}~\bibnamefont {O.}},\ }\href@noop {} {\emph {\bibinfo
  {title} {Full-Potential Electronic Structure Method}}},\ edited by\ \bibinfo
  {editor} {\bibfnamefont {E.~S.}\ \bibnamefont {H.~Dreysse}}\ and\ \bibinfo
  {editor} {\bibfnamefont {P.~P.}\ \bibnamefont {of~Solids: Springer Series~in
  Solid-State~Sciences}}\ (\bibinfo  {publisher} {Springer-Verlag},\ \bibinfo
  {address} {Berlin},\ \bibinfo {year} {2010})\BibitemShut {NoStop}%
\bibitem [{\citenamefont {Grechnev}\ \emph {et~al.}(2007)\citenamefont
  {Grechnev}, \citenamefont {Di~Marco}, \citenamefont {Katsnelson},
  \citenamefont {Lichtenstein}, \citenamefont {Wills},\ and\ \citenamefont
  {Eriksson}}]{igor}%
  \BibitemOpen
  \bibfield  {author} {\bibinfo {author} {\bibfnamefont {A.}~\bibnamefont
  {Grechnev}}, \bibinfo {author} {\bibfnamefont {I.}~\bibnamefont {Di~Marco}},
  \bibinfo {author} {\bibfnamefont {M.~I.}\ \bibnamefont {Katsnelson}},
  \bibinfo {author} {\bibfnamefont {A.~I.}\ \bibnamefont {Lichtenstein}},
  \bibinfo {author} {\bibfnamefont {J.}~\bibnamefont {Wills}}, \ and\ \bibinfo
  {author} {\bibfnamefont {O.}~\bibnamefont {Eriksson}},\ }\href {\doibase
  10.1103/PhysRevB.76.035107} {\bibfield  {journal} {\bibinfo  {journal} {Phys.
  Rev. B}\ }\textbf {\bibinfo {volume} {76}},\ \bibinfo {pages} {035107}
  (\bibinfo {year} {2007})}\BibitemShut {NoStop}%
\bibitem [{\citenamefont {Gr{\aa}n{\"a}s}\ \emph {et~al.}(2012)\citenamefont
  {Gr{\aa}n{\"a}s}, \citenamefont {Di~Marco}, \citenamefont {Thunstr{\"o}m},
  \citenamefont {Nordstr{\"o}m}, \citenamefont {Eriksson}, \citenamefont
  {Bj{\"o}rkman},\ and\ \citenamefont {Wills}}]{oscar}%
  \BibitemOpen
  \bibfield  {author} {\bibinfo {author} {\bibfnamefont {O.}~\bibnamefont
  {Gr{\aa}n{\"a}s}}, \bibinfo {author} {\bibfnamefont {I.}~\bibnamefont
  {Di~Marco}}, \bibinfo {author} {\bibfnamefont {P.}~\bibnamefont
  {Thunstr{\"o}m}}, \bibinfo {author} {\bibfnamefont {L.}~\bibnamefont
  {Nordstr{\"o}m}}, \bibinfo {author} {\bibfnamefont {O.}~\bibnamefont
  {Eriksson}}, \bibinfo {author} {\bibfnamefont {T.}~\bibnamefont
  {Bj{\"o}rkman}}, \ and\ \bibinfo {author} {\bibfnamefont {J.~M.}\
  \bibnamefont {Wills}},\ }\href {\doibase
  http://dx.doi.org/10.1016/j.commatsci.2011.11.032} {\bibfield  {journal}
  {\bibinfo  {journal} {Computational Materials Science}\ }\textbf {\bibinfo
  {volume} {55}},\ \bibinfo {pages} {295 } (\bibinfo {year}
  {2012})}\BibitemShut {NoStop}%
\bibitem [{\citenamefont {Di~Marco}\ \emph {et~al.}(2009)\citenamefont
  {Di~Marco}, \citenamefont {Min\'ar}, \citenamefont {Chadov}, \citenamefont
  {Katsnelson}, \citenamefont {Ebert},\ and\ \citenamefont
  {Lichtenstein}}]{igor-2}%
  \BibitemOpen
  \bibfield  {author} {\bibinfo {author} {\bibfnamefont {I.}~\bibnamefont
  {Di~Marco}}, \bibinfo {author} {\bibfnamefont {J.}~\bibnamefont {Min\'ar}},
  \bibinfo {author} {\bibfnamefont {S.}~\bibnamefont {Chadov}}, \bibinfo
  {author} {\bibfnamefont {M.~I.}\ \bibnamefont {Katsnelson}}, \bibinfo
  {author} {\bibfnamefont {H.}~\bibnamefont {Ebert}}, \ and\ \bibinfo {author}
  {\bibfnamefont {A.~I.}\ \bibnamefont {Lichtenstein}},\ }\href {\doibase
  10.1103/PhysRevB.79.115111} {\bibfield  {journal} {\bibinfo  {journal} {Phys.
  Rev. B}\ }\textbf {\bibinfo {volume} {79}},\ \bibinfo {pages} {115111}
  (\bibinfo {year} {2009})}\BibitemShut {NoStop}%
\bibitem [{\citenamefont {Halilov}\ \emph {et~al.}(1998)\citenamefont
  {Halilov}, \citenamefont {Eschrig}, \citenamefont {Perlov},\ and\
  \citenamefont {Oppeneer}}]{fma}%
  \BibitemOpen
  \bibfield  {author} {\bibinfo {author} {\bibfnamefont {S.~V.}\ \bibnamefont
  {Halilov}}, \bibinfo {author} {\bibfnamefont {H.}~\bibnamefont {Eschrig}},
  \bibinfo {author} {\bibfnamefont {A.~Y.}\ \bibnamefont {Perlov}}, \ and\
  \bibinfo {author} {\bibfnamefont {P.~M.}\ \bibnamefont {Oppeneer}},\ }\href
  {\doibase 10.1103/PhysRevB.58.293} {\bibfield  {journal} {\bibinfo  {journal}
  {Phys. Rev. B}\ }\textbf {\bibinfo {volume} {58}},\ \bibinfo {pages} {293}
  (\bibinfo {year} {1998})}\BibitemShut {NoStop}%
\bibitem [{\citenamefont {Liechtenstein}\ \emph {et~al.}(1987)\citenamefont
  {Liechtenstein}, \citenamefont {Katsnelson}, \citenamefont {Antropov},\ and\
  \citenamefont {Gubanov}}]{liechten1}%
  \BibitemOpen
  \bibfield  {author} {\bibinfo {author} {\bibfnamefont {A.~I.}\ \bibnamefont
  {Liechtenstein}}, \bibinfo {author} {\bibfnamefont {M.~I.}\ \bibnamefont
  {Katsnelson}}, \bibinfo {author} {\bibfnamefont {V.~P.}\ \bibnamefont
  {Antropov}}, \ and\ \bibinfo {author} {\bibfnamefont {V.~A.}\ \bibnamefont
  {Gubanov}},\ }\href {\doibase 10.1016/0304-8853(87)90721-9} {\bibfield
  {journal} {\bibinfo  {journal} {Journal of Magnetism and Magnetic Materials}\
  }\textbf {\bibinfo {volume} {67}},\ \bibinfo {pages} {65 } (\bibinfo {year}
  {1987})}\BibitemShut {NoStop}%
\bibitem [{\citenamefont {Katsnelson}\ and\ \citenamefont
  {Lichtenstein}(2000)}]{liechten2}%
  \BibitemOpen
  \bibfield  {author} {\bibinfo {author} {\bibfnamefont {M.~I.}\ \bibnamefont
  {Katsnelson}}\ and\ \bibinfo {author} {\bibfnamefont {A.~I.}\ \bibnamefont
  {Lichtenstein}},\ }\href {\doibase 10.1103/PhysRevB.61.8906} {\bibfield
  {journal} {\bibinfo  {journal} {Phys. Rev. B}\ }\textbf {\bibinfo {volume}
  {61}},\ \bibinfo {pages} {8906} (\bibinfo {year} {2000})}\BibitemShut
  {NoStop}%
\bibitem [{\citenamefont {Kvashnin}\ \emph {et~al.}(2015)\citenamefont
  {Kvashnin}, \citenamefont {Gr\aa{}n\"as}, \citenamefont {Di~Marco},
  \citenamefont {Katsnelson}, \citenamefont {Lichtenstein},\ and\ \citenamefont
  {Eriksson}}]{Jijs-in-rspt}%
  \BibitemOpen
  \bibfield  {author} {\bibinfo {author} {\bibfnamefont {Y.~O.}\ \bibnamefont
  {Kvashnin}}, \bibinfo {author} {\bibfnamefont {O.}~\bibnamefont
  {Gr\aa{}n\"as}}, \bibinfo {author} {\bibfnamefont {I.}~\bibnamefont
  {Di~Marco}}, \bibinfo {author} {\bibfnamefont {M.~I.}\ \bibnamefont
  {Katsnelson}}, \bibinfo {author} {\bibfnamefont {A.~I.}\ \bibnamefont
  {Lichtenstein}}, \ and\ \bibinfo {author} {\bibfnamefont {O.}~\bibnamefont
  {Eriksson}},\ }\href {\doibase 10.1103/PhysRevB.91.125133} {\bibfield
  {journal} {\bibinfo  {journal} {Phys. Rev. B}\ }\textbf {\bibinfo {volume}
  {91}},\ \bibinfo {pages} {125133} (\bibinfo {year} {2015})}\BibitemShut
  {NoStop}%
\bibitem [{\citenamefont {Skubic}\ \emph {et~al.}(2008)\citenamefont {Skubic},
  \citenamefont {Hellsvik}, \citenamefont {Nordstr{\"o}m},\ and\ \citenamefont
  {Eriksson}}]{uppasd}%
  \BibitemOpen
  \bibfield  {author} {\bibinfo {author} {\bibfnamefont {B.}~\bibnamefont
  {Skubic}}, \bibinfo {author} {\bibfnamefont {J.}~\bibnamefont {Hellsvik}},
  \bibinfo {author} {\bibfnamefont {L.}~\bibnamefont {Nordstr{\"o}m}}, \ and\
  \bibinfo {author} {\bibfnamefont {O.}~\bibnamefont {Eriksson}},\ }\href
  {http://stacks.iop.org/0953-8984/20/i=31/a=315203} {\bibfield  {journal}
  {\bibinfo  {journal} {Journal of Physics: Condensed Matter}\ }\textbf
  {\bibinfo {volume} {20}},\ \bibinfo {pages} {315203} (\bibinfo {year}
  {2008})}\BibitemShut {NoStop}%
\bibitem [{\citenamefont {Hwang}\ \emph {et~al.}(1996)\citenamefont {Hwang},
  \citenamefont {Cheong}, \citenamefont {Ong},\ and\ \citenamefont
  {Batlogg}}]{gtype-1}%
  \BibitemOpen
  \bibfield  {author} {\bibinfo {author} {\bibfnamefont {H.~Y.}\ \bibnamefont
  {Hwang}}, \bibinfo {author} {\bibfnamefont {S.-W.}\ \bibnamefont {Cheong}},
  \bibinfo {author} {\bibfnamefont {N.~P.}\ \bibnamefont {Ong}}, \ and\
  \bibinfo {author} {\bibfnamefont {B.}~\bibnamefont {Batlogg}},\ }\href
  {\doibase 10.1103/PhysRevLett.77.2041} {\bibfield  {journal} {\bibinfo
  {journal} {Phys. Rev. Lett.}\ }\textbf {\bibinfo {volume} {77}},\ \bibinfo
  {pages} {2041} (\bibinfo {year} {1996})}\BibitemShut {NoStop}%
\bibitem [{\citenamefont {Peng}\ \emph {et~al.}(1999)\citenamefont {Peng},
  \citenamefont {Zhao}, \citenamefont {Xie}, \citenamefont {Lin}, \citenamefont
  {Zhu}, \citenamefont {Hao}, \citenamefont {Tao}, \citenamefont {Xu},
  \citenamefont {Wang}, \citenamefont {Chen},\ and\ \citenamefont
  {Wu}}]{gtype-2}%
  \BibitemOpen
  \bibfield  {author} {\bibinfo {author} {\bibfnamefont {H.~B.}\ \bibnamefont
  {Peng}}, \bibinfo {author} {\bibfnamefont {B.~R.}\ \bibnamefont {Zhao}},
  \bibinfo {author} {\bibfnamefont {Z.}~\bibnamefont {Xie}}, \bibinfo {author}
  {\bibfnamefont {Y.}~\bibnamefont {Lin}}, \bibinfo {author} {\bibfnamefont
  {B.~Y.}\ \bibnamefont {Zhu}}, \bibinfo {author} {\bibfnamefont
  {Z.}~\bibnamefont {Hao}}, \bibinfo {author} {\bibfnamefont {H.~J.}\
  \bibnamefont {Tao}}, \bibinfo {author} {\bibfnamefont {B.}~\bibnamefont
  {Xu}}, \bibinfo {author} {\bibfnamefont {C.~Y.}\ \bibnamefont {Wang}},
  \bibinfo {author} {\bibfnamefont {H.}~\bibnamefont {Chen}}, \ and\ \bibinfo
  {author} {\bibfnamefont {F.}~\bibnamefont {Wu}},\ }\href {\doibase
  10.1103/PhysRevLett.82.362} {\bibfield  {journal} {\bibinfo  {journal} {Phys.
  Rev. Lett.}\ }\textbf {\bibinfo {volume} {82}},\ \bibinfo {pages} {362}
  (\bibinfo {year} {1999})}\BibitemShut {NoStop}%
\bibitem [{\citenamefont {Zaanen}\ \emph {et~al.}(1985)\citenamefont {Zaanen},
  \citenamefont {Sawatzky},\ and\ \citenamefont {Allen}}]{mott-ins}%
  \BibitemOpen
  \bibfield  {author} {\bibinfo {author} {\bibfnamefont {J.}~\bibnamefont
  {Zaanen}}, \bibinfo {author} {\bibfnamefont {G.~A.}\ \bibnamefont
  {Sawatzky}}, \ and\ \bibinfo {author} {\bibfnamefont {J.~W.}\ \bibnamefont
  {Allen}},\ }\href {\doibase 10.1103/PhysRevLett.55.418} {\bibfield  {journal}
  {\bibinfo  {journal} {Phys. Rev. Lett.}\ }\textbf {\bibinfo {volume} {55}},\
  \bibinfo {pages} {418} (\bibinfo {year} {1985})}\BibitemShut {NoStop}%
\bibitem [{\citenamefont {Nguyen}\ \emph {et~al.}(2011)\citenamefont {Nguyen},
  \citenamefont {Bach}, \citenamefont {Pham}, \citenamefont {Pham},
  \citenamefont {Nguyen},\ and\ \citenamefont {Hoang}}]{nguyen}%
  \BibitemOpen
  \bibfield  {author} {\bibinfo {author} {\bibfnamefont {T.~T.}\ \bibnamefont
  {Nguyen}}, \bibinfo {author} {\bibfnamefont {T.~C.}\ \bibnamefont {Bach}},
  \bibinfo {author} {\bibfnamefont {H.~T.}\ \bibnamefont {Pham}}, \bibinfo
  {author} {\bibfnamefont {T.~T.}\ \bibnamefont {Pham}}, \bibinfo {author}
  {\bibfnamefont {D.~T.}\ \bibnamefont {Nguyen}}, \ and\ \bibinfo {author}
  {\bibfnamefont {N.~N.}\ \bibnamefont {Hoang}},\ }\href {\doibase
  http://dx.doi.org/10.1016/j.physb.2011.06.054} {\bibfield  {journal}
  {\bibinfo  {journal} {Physica B: Condensed Matter}\ }\textbf {\bibinfo
  {volume} {406}},\ \bibinfo {pages} {3613 } (\bibinfo {year}
  {2011})}\BibitemShut {NoStop}%
\bibitem [{\citenamefont {Cardoso}\ \emph {et~al.}(2008)\citenamefont
  {Cardoso}, \citenamefont {Borges}, \citenamefont {Gasche},\ and\
  \citenamefont {Godinho}}]{cardoso}%
  \BibitemOpen
  \bibfield  {author} {\bibinfo {author} {\bibfnamefont {C.}~\bibnamefont
  {Cardoso}}, \bibinfo {author} {\bibfnamefont {R.~P.}\ \bibnamefont {Borges}},
  \bibinfo {author} {\bibfnamefont {T.}~\bibnamefont {Gasche}}, \ and\ \bibinfo
  {author} {\bibfnamefont {M.}~\bibnamefont {Godinho}},\ }\href
  {http://stacks.iop.org/0953-8984/20/i=3/a=035202} {\bibfield  {journal}
  {\bibinfo  {journal} {Journal of Physics: Condensed Matter}\ }\textbf
  {\bibinfo {volume} {20}},\ \bibinfo {pages} {035202} (\bibinfo {year}
  {2008})}\BibitemShut {NoStop}%
\bibitem [{\citenamefont {Domb}\ \emph {et~al.}(1984)\citenamefont {Domb},
  \citenamefont {Green},\ and\ \citenamefont {Lebowitz}}]{j-exp}%
  \BibitemOpen
  \bibfield  {author} {\bibinfo {author} {\bibfnamefont {C.}~\bibnamefont
  {Domb}}, \bibinfo {author} {\bibfnamefont {M.}~\bibnamefont {Green}}, \ and\
  \bibinfo {author} {\bibfnamefont {J.}~\bibnamefont {Lebowitz}},\ }\href
  {https://books.google.se/books?id=3pIpAQAAMAAJ} {\emph {\bibinfo {title}
  {Phase Transitions and Critical Phenomena}}},\ \bibinfo {series} {Phase
  Transitions and Critical Phenomena}\ No.\ \bibinfo {number} {v. 9}\ (\bibinfo
   {publisher} {Academic Press},\ \bibinfo {year} {1984})\BibitemShut {NoStop}%
\bibitem [{\citenamefont {Mazin}(2007)}]{mazin}%
  \BibitemOpen
  \bibfield  {author} {\bibinfo {author} {\bibfnamefont {I.~I.}\ \bibnamefont
  {Mazin}},\ }\href {\doibase 10.1103/PhysRevB.75.094407} {\bibfield  {journal}
  {\bibinfo  {journal} {Phys. Rev. B}\ }\textbf {\bibinfo {volume} {75}},\
  \bibinfo {pages} {094407} (\bibinfo {year} {2007})}\BibitemShut {NoStop}%
\bibitem [{\citenamefont {Anisimov}\ \emph {et~al.}(1991)\citenamefont
  {Anisimov}, \citenamefont {Zaanen},\ and\ \citenamefont
  {Andersen}}]{Hubbard-Stoner}%
  \BibitemOpen
  \bibfield  {author} {\bibinfo {author} {\bibfnamefont {V.~I.}\ \bibnamefont
  {Anisimov}}, \bibinfo {author} {\bibfnamefont {J.}~\bibnamefont {Zaanen}}, \
  and\ \bibinfo {author} {\bibfnamefont {O.~K.}\ \bibnamefont {Andersen}},\
  }\href {\doibase 10.1103/PhysRevB.44.943} {\bibfield  {journal} {\bibinfo
  {journal} {Phys. Rev. B}\ }\textbf {\bibinfo {volume} {44}},\ \bibinfo
  {pages} {943} (\bibinfo {year} {1991})}\BibitemShut {NoStop}%
\bibitem [{\citenamefont {Solovyev}\ and\ \citenamefont
  {Terakura}(1999)}]{super-double}%
  \BibitemOpen
  \bibfield  {author} {\bibinfo {author} {\bibfnamefont {I.~V.}\ \bibnamefont
  {Solovyev}}\ and\ \bibinfo {author} {\bibfnamefont {K.}~\bibnamefont
  {Terakura}},\ }\href {\doibase 10.1103/PhysRevLett.82.2959} {\bibfield
  {journal} {\bibinfo  {journal} {Phys. Rev. Lett.}\ }\textbf {\bibinfo
  {volume} {82}},\ \bibinfo {pages} {2959} (\bibinfo {year}
  {1999})}\BibitemShut {NoStop}%
\bibitem [{\citenamefont {Azumi}\ and\ \citenamefont {Goldman}(1954)}]{bethe}%
  \BibitemOpen
  \bibfield  {author} {\bibinfo {author} {\bibfnamefont {K.}~\bibnamefont
  {Azumi}}\ and\ \bibinfo {author} {\bibfnamefont {J.~E.}\ \bibnamefont
  {Goldman}},\ }\href {\doibase 10.1103/PhysRev.93.630} {\bibfield  {journal}
  {\bibinfo  {journal} {Phys. Rev.}\ }\textbf {\bibinfo {volume} {93}},\
  \bibinfo {pages} {630} (\bibinfo {year} {1954})}\BibitemShut {NoStop}%
\bibitem [{\citenamefont {Autieri}(2016)}]{carmine}%
  \BibitemOpen
  \bibfield  {author} {\bibinfo {author} {\bibfnamefont {C.}~\bibnamefont
  {Autieri}},\ }\href {http://stacks.iop.org/0953-8984/28/i=42/a=426004}
  {\bibfield  {journal} {\bibinfo  {journal} {Journal of Physics: Condensed
  Matter}\ }\textbf {\bibinfo {volume} {28}},\ \bibinfo {pages} {426004}
  (\bibinfo {year} {2016})}\BibitemShut {NoStop}%
\end{thebibliography}

%

\end{document}